\documentclass{aastex63}

\usepackage{hyperref}
\usepackage{graphicx}
\usepackage{here}
\usepackage{lineno}
\usepackage{multirow}

\newcommand{\average}[1]{\ensuremath{\langle #1 \rangle} }
\usepackage{mhchem}
\usepackage{isotope}

\begin{document}

\title{Search for supernova bursts in Super-Kamiokande IV}
\newcommand{\AFFicrr}{\affiliation{Kamioka Observatory, Institute for Cosmic Ray Research, University of Tokyo, Kamioka, Gifu 506-1205, Japan}}
\newcommand{\AFFkashiwa}{\affiliation{Research Center for Cosmic Neutrinos, Institute for Cosmic Ray Research, University of Tokyo, Kashiwa, Chiba 277-8582, Japan}}
\newcommand{\AFFicrronly}{\affiliation{Institute for Cosmic Ray Research, University of Tokyo, Kashiwa, Chiba 277-8582, Japan}}
\newcommand{\AFFipmu}{\affiliation{Kavli Institute for the Physics and
Mathematics of the Universe (WPI), The University of Tokyo Institutes for Advanced Study,
University of Tokyo, Kashiwa, Chiba 277-8583, Japan }}
\newcommand{\AFFmad}{\affiliation{Department of Theoretical Physics, University Autonoma Madrid, 28049 Madrid, Spain}}
\newcommand{\AFFubc}{\affiliation{Department of Physics and Astronomy, University of British Columbia, Vancouver, BC, V6T1Z4, Canada}}
\newcommand{\AFFbu}{\affiliation{Department of Physics, Boston University, Boston, MA 02215, USA}}
\newcommand{\AFFuci}{\affiliation{Department of Physics and Astronomy, University of California, Irvine, Irvine, CA 92697-4575, USA }}
\newcommand{\AFFcsu}{\affiliation{Department of Physics, California State University, Dominguez Hills, Carson, CA 90747, USA}}
\newcommand{\AFFcnm}{\affiliation{Institute for Universe and Elementary Particles, Chonnam National University, Gwangju 61186, Korea}}
\newcommand{\AFFduke}{\affiliation{Department of Physics, Duke University, Durham NC 27708, USA}}
\newcommand{\AFFfukuoka}{\affiliation{Junior College, Fukuoka Institute of Technology, Fukuoka, Fukuoka 811-0295, Japan}}
\newcommand{\AFFgifu}{\affiliation{Department of Physics, Gifu University, Gifu, Gifu 501-1193, Japan}}
\newcommand{\AFFgist}{\affiliation{GIST College, Gwangju Institute of Science and Technology, Gwangju 500-712, Korea}}
\newcommand{\AFFuh}{\affiliation{Department of Physics and Astronomy, University of Hawaii, Honolulu, HI 96822, USA}}
\newcommand{\AFFicl}{\affiliation{Department of Physics, Imperial College London , London, SW7 2AZ, United Kingdom }}
\newcommand{\AFFkek}{\affiliation{High Energy Accelerator Research Organization (KEK), Tsukuba, Ibaraki 305-0801, Japan }}
\newcommand{\AFFkobe}{\affiliation{Department of Physics, Kobe University, Kobe, Hyogo 657-8501, Japan}}
\newcommand{\AFFkyoto}{\affiliation{Department of Physics, Kyoto University, Kyoto, Kyoto 606-8502, Japan}}
\newcommand{\AFFliv}{\affiliation{Department of Physics, University of Liverpool, Liverpool, L69 7ZE, United Kingdom}}
\newcommand{\AFFmiyagi}{\affiliation{Department of Physics, Miyagi University of Education, Sendai, Miyagi 980-0845, Japan}}
\newcommand{\AFFnagoya}{\affiliation{Institute for Space-Earth Environmental Research, Nagoya University, Nagoya, Aichi 464-8602, Japan}}
\newcommand{\AFFkmi}{\affiliation{Kobayashi-Maskawa Institute for the Origin of Particles and the Universe, Nagoya University, Nagoya, Aichi 464-8602, Japan}}
\newcommand{\AFFpol}{\affiliation{National Centre For Nuclear Research, 02-093 Warsaw, Poland}}
\newcommand{\AFFsuny}{\affiliation{Department of Physics and Astronomy, State University of New York at Stony Brook, NY 11794-3800, USA}}
\newcommand{\AFFokayama}{\affiliation{Department of Physics, Okayama University, Okayama, Okayama 700-8530, Japan }}
\newcommand{\AFFosaka}{\affiliation{Department of Physics, Osaka University, Toyonaka, Osaka 560-0043, Japan}}
\newcommand{\AFFox}{\affiliation{Department of Physics, Oxford University, Oxford, OX1 3PU, United Kingdom}}
\newcommand{\AFFqmul}{\affiliation{School of Physics and Astronomy, Queen Mary University of London, London, E1 4NS, United Kingdom}}
\newcommand{\AFFregina}{\affiliation{Department of Physics, University of Regina, 3737 Wascana Parkway, Regina, SK, S4SOA2, Canada}}
\newcommand{\AFFseoul}{\affiliation{Department of Physics, Seoul National University, Seoul 151-742, Korea}}
\newcommand{\AFFsheff}{\affiliation{Department of Physics and Astronomy, University of Sheffield, S3 7RH, Sheffield, United Kingdom}}
\newcommand{\AFFshizuokasc}{\affiliation{Department of Informatics in
Social Welfare, Shizuoka University of Welfare, Yaizu, Shizuoka, 425-8611, Japan}}
\newcommand{\AFFstfc}{\affiliation{STFC, Rutherford Appleton Laboratory, Harwell Oxford, and Daresbury Laboratory, Warrington, OX11 0QX, United Kingdom}}
\newcommand{\AFFskk}{\affiliation{Department of Physics, Sungkyunkwan University, Suwon 440-746, Korea}}
\newcommand{\AFFtokyo}{\affiliation{The University of Tokyo, Bunkyo, Tokyo 113-0033, Japan }}
\newcommand{\AFFtodai}{\affiliation{Department of Physics, University of Tokyo, Bunkyo, Tokyo 113-0033, Japan }}
\newcommand{\AFFtit}{\affiliation{Department of Physics,Tokyo Institute of Technology, Meguro, Tokyo 152-8551, Japan }}
\newcommand{\AFFtus}{\affiliation{Department of Physics, Faculty of Science and Technology, Tokyo University of Science, Noda, Chiba 278-8510, Japan }}
\newcommand{\AFFtoronto}{\affiliation{Department of Physics, University of Toronto, ON, M5S 1A7, Canada }}
\newcommand{\AFFtriumf}{\affiliation{TRIUMF, 4004 Wesbrook Mall, Vancouver, BC, V6T2A3, Canada }}
\newcommand{\AFFtokai}{\affiliation{Department of Physics, Tokai University, Hiratsuka, Kanagawa 259-1292, Japan}}
\newcommand{\AFFtsinghua}{\affiliation{Department of Engineering Physics, Tsinghua University, Beijing, 100084, China}}
\newcommand{\AFFynu}{\affiliation{Department of Physics, Yokohama National University, Yokohama, Kanagawa, 240-8501, Japan}}
\newcommand{\AFFllr}{\affiliation{Ecole Polytechnique, IN2P3-CNRS, Laboratoire Leprince-Ringuet, F-91120 Palaiseau, France }}
\newcommand{\AFFbari}{\affiliation{ Dipartimento Interuniversitario di Fisica, INFN Sezione di Bari and Universit\`a e Politecnico di Bari, I-70125, Bari, Italy}}
\newcommand{\AFFnapoli}{\affiliation{Dipartimento di Fisica, INFN Sezione di Napoli and Universit\`a di Napoli, I-80126, Napoli, Italy}}
\newcommand{\AFFroma}{\affiliation{INFN Sezione di Roma and Universit\`a di Roma ``La Sapienza'', I-00185, Roma, Italy}}
\newcommand{\AFFpadova}{\affiliation{Dipartimento di Fisica, INFN Sezione di Padova and Universit\`a di Padova, I-35131, Padova, Italy}}
\newcommand{\AFFkeio}{\affiliation{Department of Physics, Keio University, Yokohama, Kanagawa, 223-8522, Japan}}
\newcommand{\AFFwinnipeg}{\affiliation{Department of Physics, University of Winnipeg, MB R3J 3L8, Canada }}
\newcommand{\AFFkcl}{\affiliation{Department of Physics, King's College London, London, WC2R 2LS, UK }}
\newcommand{\AFFwarwick}{\affiliation{Department of Physics, University of Warwick, Coventry, CV4 7AL, UK }}
\newcommand{\AFFral}{\affiliation{Rutherford Appleton Laboratory, Harwell, Oxford, OX11 0QX, UK }}
\newcommand{\AFFwu}{\affiliation{Faculty of Physics, University of Warsaw, Warsaw, 02-093, Poland }}
\newcommand{\AFFbcit}{\affiliation{Department of Physics, British Columbia Institute of Technology, Burnaby, BC, V5G 3H2, Canada }}
\newcommand{\AFFifirse}{\affiliation{Institute For Interdisciplinary Research in Science and Education, ICISE, Quy Nhon, 55121, Vietnam}}
\newcommand{\AFFtohoku}{\affiliation{Department of Physics, Faculty of Science, Tohoku University, Sendai, Miyagi, 980-8578, Japan}}
\newcommand{\AFFilance}{\affiliation{International Laboratory for Astrophysics, Neutrino and Cosmology Experiment, Kashiwa, Chiba 277-8582, Japan}}
\newcommand{\AFFibs}{\affiliation{Institute for Basic Science (IBS), Daejeon, 34126, Korea}}

\AFFicrr
\AFFkashiwa
\AFFicrronly
\AFFmad
\AFFbu
\AFFbcit
\AFFuci
\AFFcsu
\AFFcnm
\AFFduke
\AFFllr
\AFFfukuoka
\AFFgifu
\AFFgist
\AFFuh
\AFFicl
\AFFbari
\AFFnapoli
\AFFpadova
\AFFroma
\AFFilance
\AFFkeio
\AFFkek
\AFFkcl
\AFFkobe
\AFFkyoto
\AFFliv
\AFFmiyagi
\AFFnagoya
\AFFkmi
\AFFpol
\AFFsuny
\AFFokayama
\AFFox
\AFFral
\AFFseoul
\AFFsheff
\AFFshizuokasc
\AFFstfc
\AFFskk
\AFFtokai
\AFFtokyo
\AFFtodai
\AFFipmu
\AFFtit
\AFFtus
\AFFtoronto
\AFFtriumf
\AFFtsinghua
\AFFwu
\AFFwarwick
\AFFwinnipeg
\AFFynu
\AFFibs

\author{M.~Mori}
\AFFkyoto

\author{K.~Abe}
\AFFicrr
\AFFipmu
\author{Y.~Hayato}
\AFFicrr
\AFFipmu
\author{K.~Hiraide}
\AFFicrr
\AFFipmu
\author{K.~Ieki}
\AFFicrr
\author{M.~Ikeda}
\author{S.~Imaizumi}
\AFFicrr
\author{J.~Kameda}
\AFFicrr
\AFFipmu
\author{Y.~Kanemura}
\AFFicrr
\author{R.~Kaneshima}
\AFFicrr
% ...

\collaboration{246}{The Super-Kamiokande Collaboration}
\noaffiliation


\begin{abstract}
Super-Kamiokande has been searching for neutrino bursts characteristic of core-collapse supernovae continuously, in real time, since the start of operations in 1996. The present work focuses on detecting more distant supernovae whose event rate may be too small to trigger in real time, but may be identified using an offline approach. The analysis of data collected from 2008 to 2018 found no evidence of distant supernovae bursts. This establishes an upper limit of 0.29 year$^{-1}$ on the rate of core-collapse supernovae out to 100 kpc at 90\% C.L..
For supernovae that fail to explode and collapse directly to black holes the limit reaches to 300 kpc.
\end{abstract}
\vspace{10pt}
%
%
%
%
%
\section{Introduction}\label{sec:intro}
The observation of neutrinos from SN1987A~\citep{PhysRevLett.58.1490,PhysRevLett.58.1494,ALEXEYEV1988209,AGLIETTA1988453} established the basic mechanism of core-collapse supernovae\footnote{Note that there are type Ia supernovae that are driven by other mechanisms and do not emit many neutrinos~\cite{Wright_2016}. In this paper we use the term "supernova" to mean core-collapse supernova.} 
featuring neutrino production during the initial matter infall, the subsequent shock revival and at later times as the remnant proto-neutron star cools~\citep{PhysRevLett.58.2722}. This production is expected to last several tens of seconds and to carry detailed information on the mechanisms and dynamics of the collapse process.

Since that first observation, several experiments have searched in real time for other galactic supernovae
using liquid scintillator~\citep{Asakura:2015bga,Bruno:2017oxg,Monzani:2006jg}, plastic scintillator~\citep{MACRO:1997enu}, lead~\citep{Zuber:2015ita}, heavy-water~\citep{Aharmim_2011}, and water-based detectors~\citep{Abe:2016waf, Kopke:2017req}. These have yielded only null results. Such searches focus on the intense burst of neutrinos expected for a canonical explosion at the galactic center (about 10~kpc) in order to overcome intrinsic backgrounds and trigger the detector, as in \citep{Ikeda:2007sa}. More distant supernovae likely only produce a weak signal incapable of being detected by such searches. However, this signal has the characteristic distributions of neutrino interactions in space and time that may be used to extract a viable signal from the backgrounds. Such events are the target of the present search. 

Although distant supernovae produce only a few neutrino interactions in a terrestrial detector  
and provide limited information about the collapse mechanism, their observation has other advantages. 
For example, the accumulation of stars and dust along the galactic plane as viewed from Earth 
obfuscates optical observations of supernova occurring opposite our position from the galactic center 
but is no barrier to the neutrino signal. As well, some stars may collapse as so-called ``failed supernova'', emitting neutrinos without a corresponding optical counterpart~\citep{Nakazato:2012qf} due to the formation of a black hole. 

Off-line searches have the potential to uncover hitherto unseen supernovae or to constrain their rate. 
Indeed, present estimates of the rate of supernovae from observational data are $O(1)$ per 100 years per galaxy and are accompanied by large uncertainties~\citep{Rozwadowska:2021lll}. Note that based on optical searches~\citep{PhysRevLett.95.171101}, the supernova rate out to 60~kpc is estimated to be about 0.035~year$^{-1}$.
As there are few known stars between 60~kpc and 100~kpc, there is no rate expected in this region. 
However, a neutrino search allows for constraints on the opposite side of the galaxy where optical searches are limited and the detection of a positive signal with only a few neutrino interactions would nevertheless allow for comparison against models. 
Indeed, the detection of only two dozen neutrinos from SN1987A yielded useful gross variables such as the total neutrino luminosity and the average neutrino energy.  
Since then distant supernovae searches with neutrinos have have been performed at other experiments, including the 1 kton liquid scintillator detector LVD~\citep{Bruno:2017oxg,Vigorito:2020vjj} and at Super-Kamiokande using earlier data~\citep{Ikeda:2007sa}.

This paper describes the search for distant supernova in Super-Kamiokande data from the period 2008 to 2018, using an updated analysis technique and newer supernovae neutrino emission models. Sections~\ref{sec:sk}
and~\ref{sec:methodology} 
present the detector and the search methodology, respectively. 
A description of the analysis sensitivity is presented in
Section~\ref{sec:performance} 
before the search results are described in 
Section~\ref{sec:result}. Discussion of the result is described in Section~\ref{sec:discussion}.
Finally, concluding remarks are made in 
Section~\ref{sec:summary}.

\section{The Super-Kamiokande Experiment}\label{sec:sk}
Super-Kamiokande(SK) is a water Cherenkov detector located $\sim$ 1000~m deep under Mt. Ikeno in the Gifu prefecture, Japan. 
Its 50 kilotons of ultra pure water are held in a cylindrical stainless steel tank that is optically separated into two regions~\citep{Fukuda:2002uc}. 
The inner detector (ID) forms the main target volume. 
For the data period in this work it was instrumented with 11,129 20-inch photomultiplier tubes (PMTs) that observe Cherenkov light from charged particles traversing its 32~kton water volume.
The outer detector (OD) serves primarily as a veto of external radiation in the 18~kton water volume surrounding the ID and is instrumented with 8-inch PMTs connected to wavelength-shifting plates. 
Due to radioactive backgrounds emitted from detector materials in the PMT support structure, most analyses 
use a 22.5~kton ``fiducial'' sub-volume of the ID (or smaller volumes at low energies)~\citep{Abe:2016nxk}.
However, in the event of a nearby galactic supernova, the neutrino event rate is expected to overwhelm 
those backgrounds such that the entire ID volume can be used. 
As discussed below, the present analysis will make use of both the fiducial and full ID volumes in a staged approach to identifying and 
verifying a distant supernova signal.

SK has been in operation since April 1996 and has spanned six run periods to date, labeled SK-I through SK-VI. 
During the SK-I period the detector was operated with 40\% photocathode coverage until July 2001, when it was stopped for maintenance.
Following a subsequent accident that destroyed half of the PMTs, SK-II took data with 19\% coverage from 2002 until 2005. 
After recovering the full complement of PMTs in 2006 the SK-III period continued until 2008 when it was stopped
for one week to upgrade the front-end electronics~\citep{5446533}. 
During this upgrade the electronics were replaced in sections so as to minimize the number of inactive channels at any give time and to minimize the dead time for supernovae detection.
The subsequent period, SK-IV, ran for the next ten years. 
In the summer of 2018 the detector was again stopped for maintenance and repairs in preparation for the addition of 
gadolinium sulfate to the detector water~\citep{Super-Kamiokande:2021the}. 
The final pure water phase, SK-V, was operated from 2019 until summer of 2020, when the gadolinium compound 
was added for the first time, marking the start of the SK-Gd project and SK-VI.

This analysis uses the full SK-IV data set, corresponding to 3381.41 live-days. 
In contrast to earlier phases and the previous distant supernova search, 
the SK-IV period is characterized by dead-time-free electronics using a low noise charge-to-time converter (QTC) design~\citep{Nishino:2009zu} which has enabled a lower energy threshold. 
The event trigger used in this analysis is formed when at least 47 PMTs register hits within any 200 nanosecond time period.
This corresponds to efficiency of 54\% for positron kinetic energies greater than 5.5~MeV and 99\% for energies greater than 8.5~MeV.
Although lower energy triggers are available, the majority of supernova neutrino candidates are expected to be above 5~MeV. In this paper, the term ``event'' refers to an interaction within the detector satisfying this criterion.


\section{Search Methodology}\label{sec:methodology}
Neutrinos from supernovae are expected to produce a collection of interactions (a ``cluster'') in a short time interval with a uniform spatial distribution in a terrestrial detector. 
For a supernova near the galactic center (10~kpc) such a cluster would contain O(1000) events in SK distributed over 100 seconds, with the majority occurring in the first 10 seconds. 
The number of events in a cluster decreases with the inverse square of the distance.

In order to search for these distant supernovae, this new analysis expands the methods in a previous search ~\citep{Ikeda:2007sa} and the methods used in the SK real-time supernova monitor system, ``SNWatch''~\citep{Abe:2016waf}. 
The new method first searches for clusters of events in the 22.5~kton fiducial volume using three different combinations of energy thresholds and time windows, the longest of which is 10~s (described below). 
Clusters are then subject to additional selection cuts to remove backgrounds stemming from cosmic-muon-induced spallation or electronic noise. These cuts have been optimized using recent supernova neutrino models, described below. If a cluster survives the cuts, the search for additional cluster events will be expanded to include the whole ID volume near that cluster's time to either bolster confidence in the signal or reject rare backgrounds that have a signal-like topology. 

\subsection{Supernova models}\label{sec:supernova_model}
\begin{figure}[h]
    \centering
    \begin{tabular}{cc}
         \begin{minipage}{0.5\hsize}
         \centering
         \includegraphics[width=8cm]{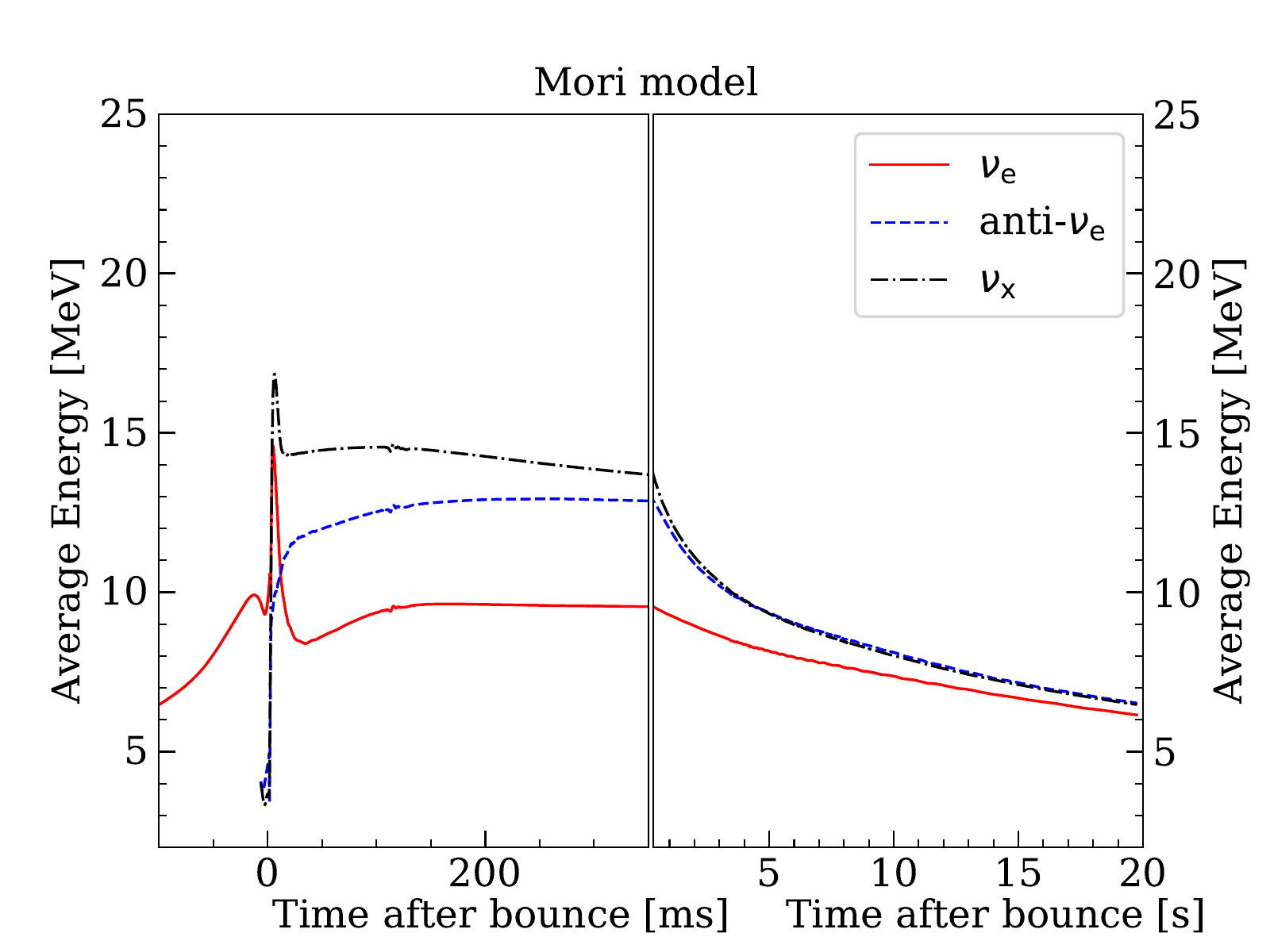}
         \end{minipage}&
         \begin{minipage}{0.5\hsize} 
         \centering
         \includegraphics[keepaspectratio, width=8cm]{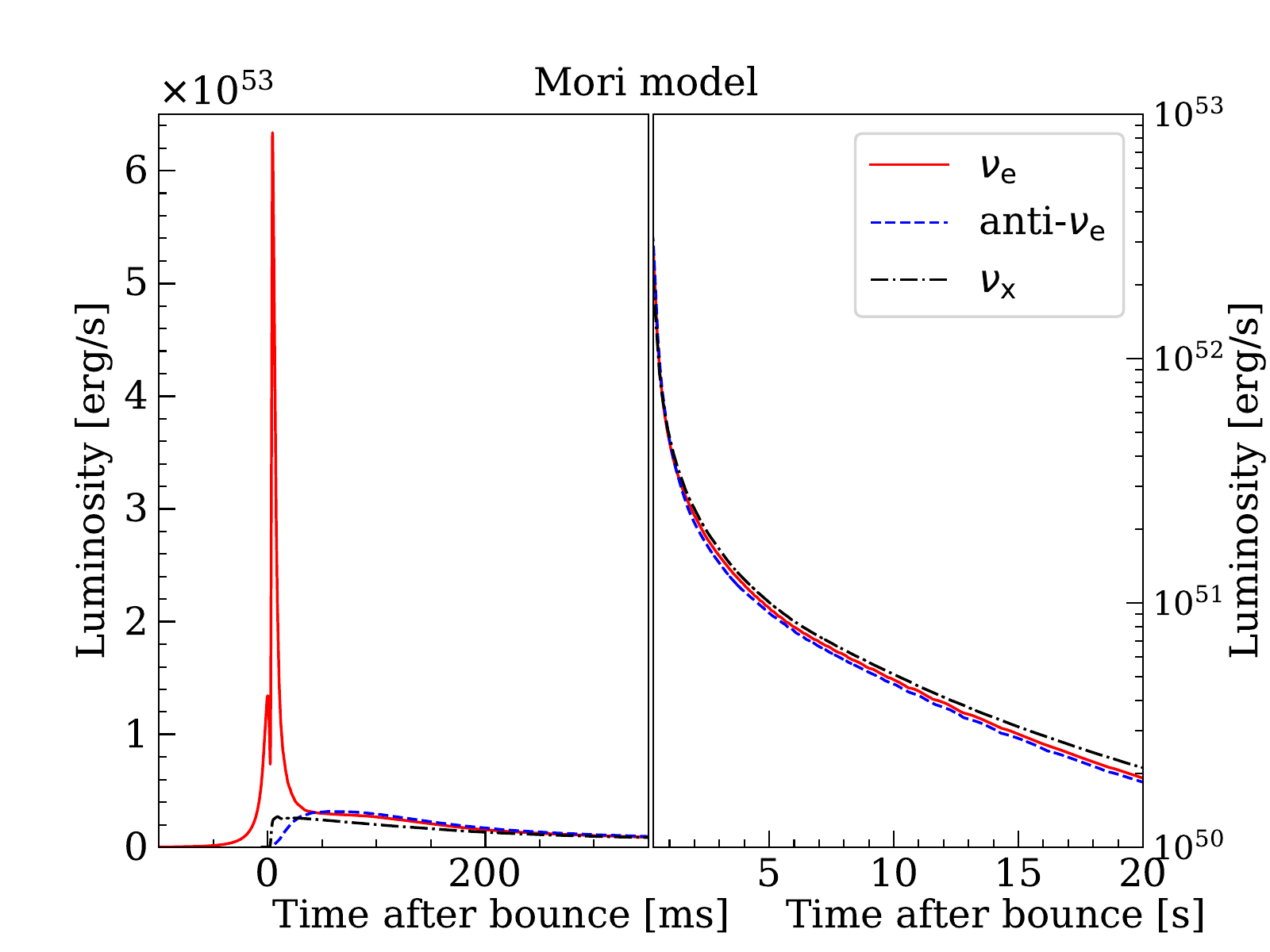}
         \end{minipage} \\
        \begin{minipage}{0.5\hsize}
         \centering
        \includegraphics[keepaspectratio, width=8cm]{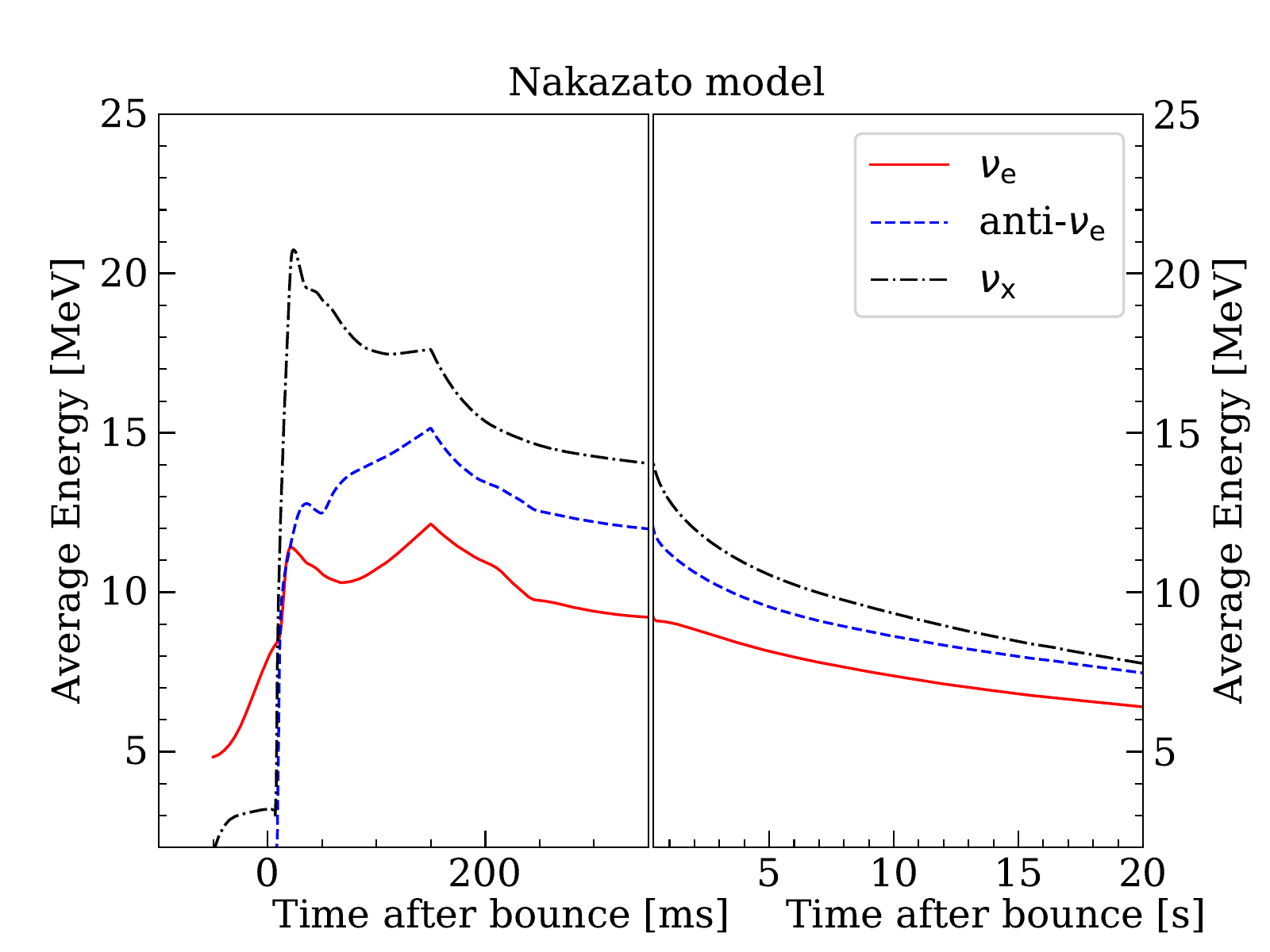}
         \end{minipage}&
         \begin{minipage}{0.5\hsize}
         \centering
         \includegraphics[keepaspectratio, width=8cm]{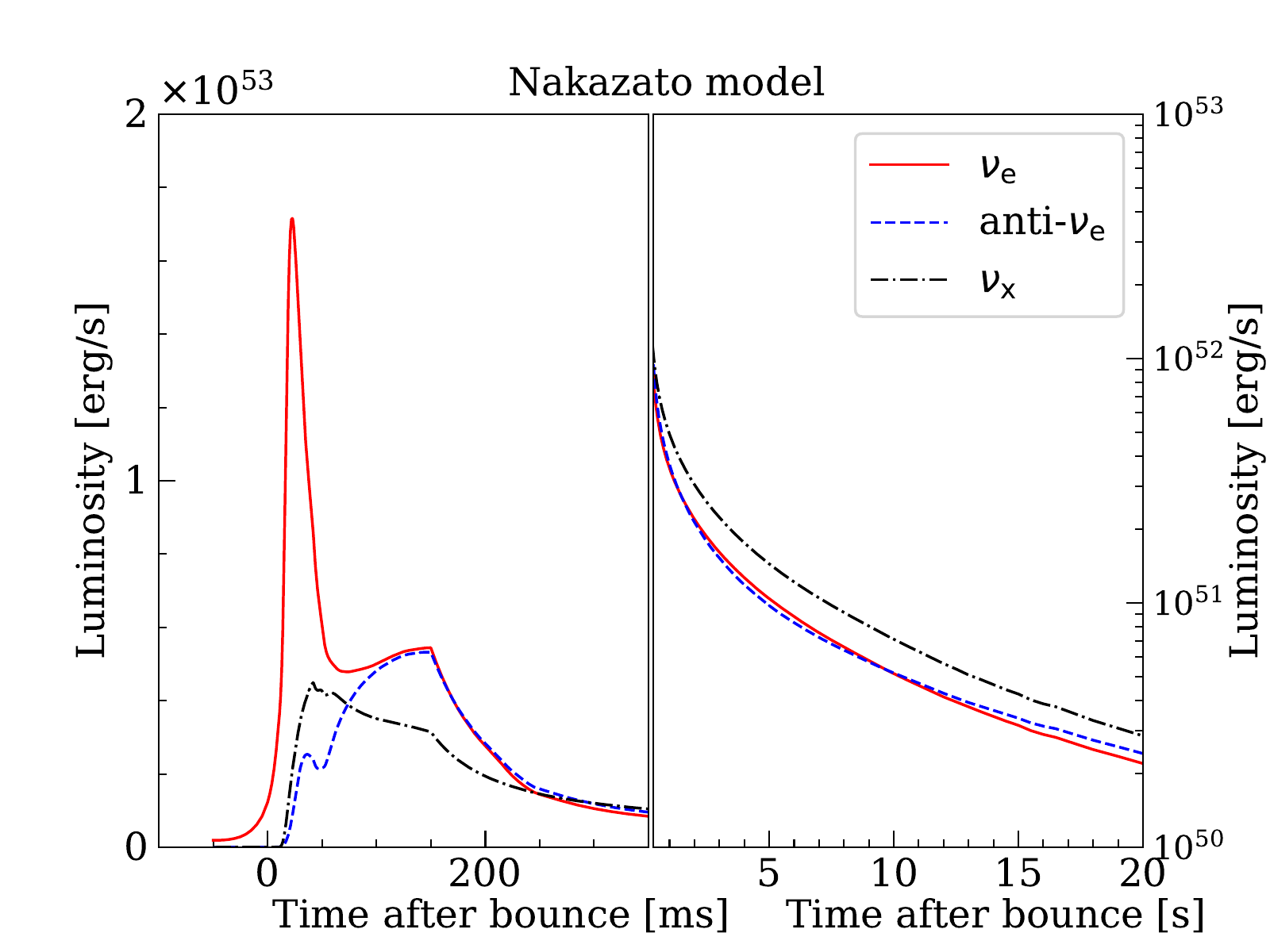}
         \end{minipage} \\
        \begin{minipage}{0.5\hsize}
         \centering
        \includegraphics[keepaspectratio, width=8cm]{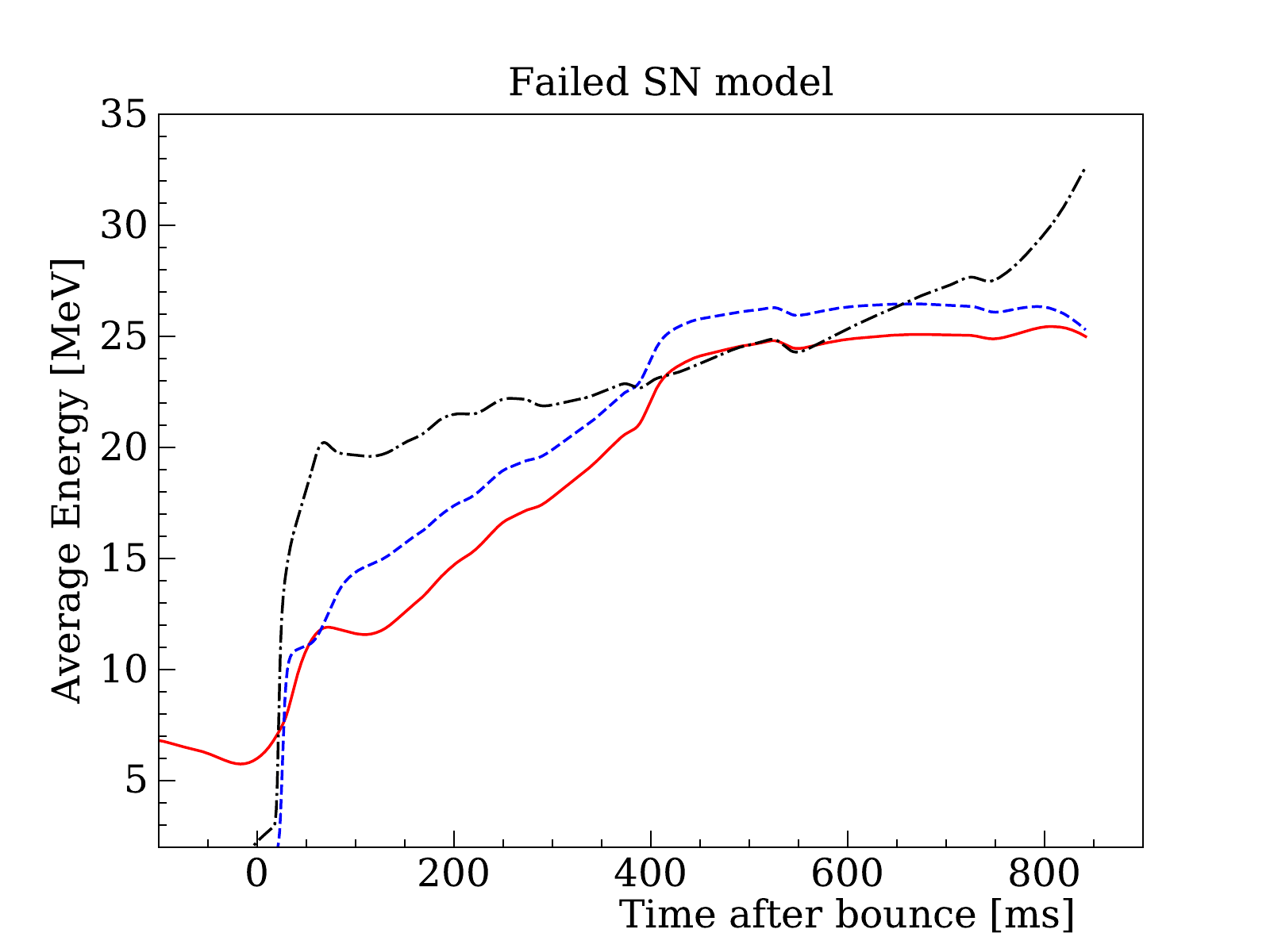}
         \end{minipage}&
         \begin{minipage}{0.5\hsize}
         \centering
         \includegraphics[keepaspectratio, width=8cm]{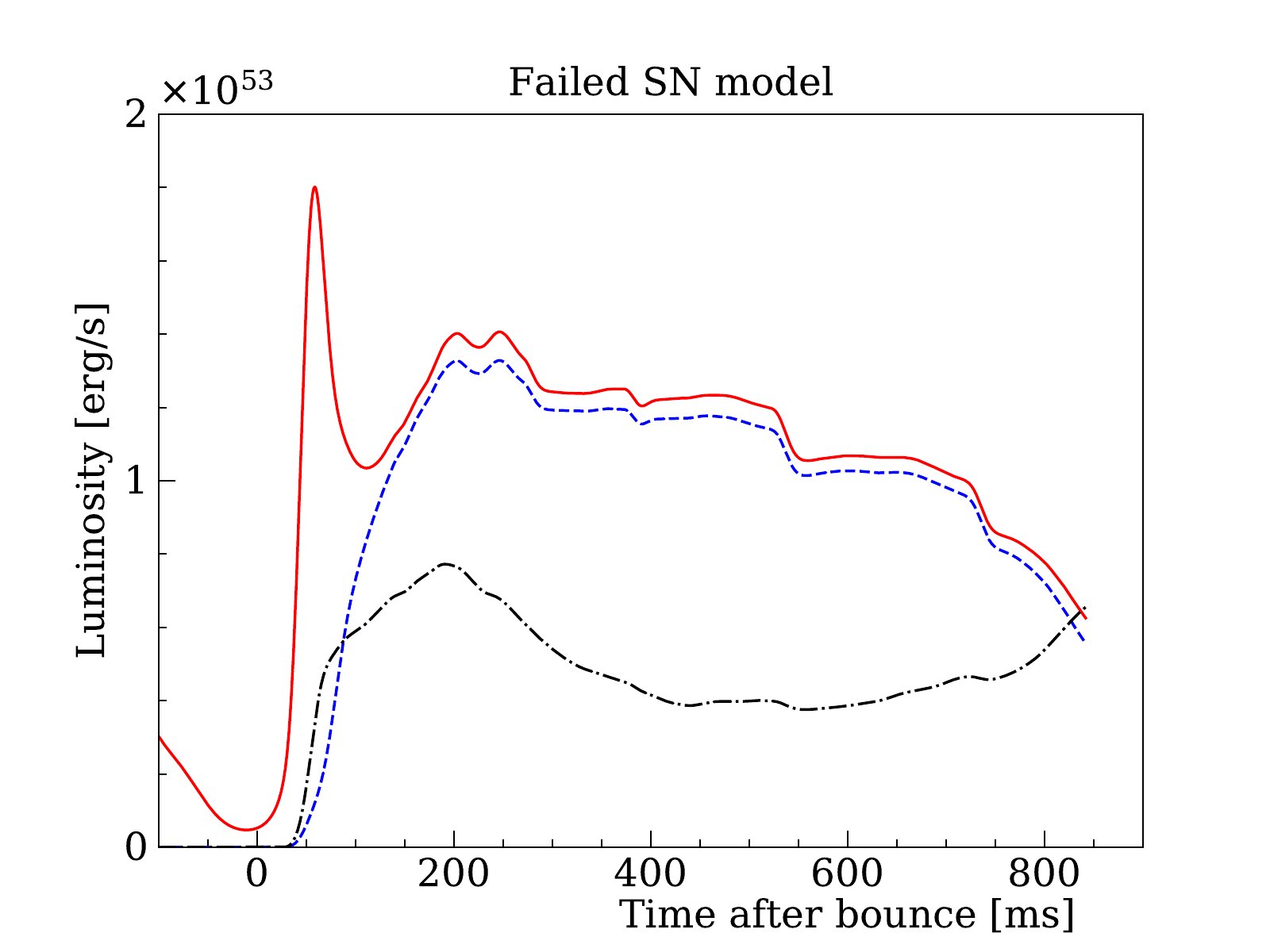}
         \end{minipage}
    \end{tabular}
    \caption{Average energy and luminosity of neutrinos in the Mori (top), Nakazato (middle) and failed SN (bottom) models. The left panels show the 
average energy and the right panels show the luminosity.
The horizontal axis shows the time relative to the supernova bounce, where the figures are divided into the 
region -100~ms to 350~ms and 350~ms to 20~s for the Mori and Nakazato models. The horizontal axis for the failed SN model ranges up to 800~ms after the bounce.
Red lines denote electron neutrinos, blue lines their antineutrinos, and the black lines represent all other neutrino species.}
\label{fig:supernovamodels}
\end{figure}

Since supernovae producing a high-intensity neutrino burst in the detector are likely to have been identified by the real-time supernova monitor, this search focuses on clusters with a small number of events produced within up to 10 seconds. 
While many supernova models offer predictions of neutrino emission for only the first second of the burst, the Livermore model~\citep{totani} provides predictions up to 20 seconds. Until recently the Livermore model was the only simulation with predictions out to such long times. This late time property of the model was used in the previous search using earlier SK data~\citep{Ikeda:2007sa}.
However, the present analysis adopts three newer models: the Mori model~\citep{10.1093/ptep/ptaa185}, the Nakazato model~\citep{Nakazato:2012qf} and a model of failed SN~\citep{Nakazato:2012qf}. Each model has more advanced treatments of relativity, neutrino transport, and neutrino interactions during the burst. The former two models are spherically symmetric supernova simulations that explode successfully. The latter model is also one-dimensional, but the star fails to explode and instead produces a black hole. These models address other known issues in the Livermore model, such as the rise in the average neutrino energy at late times.
In this study, the Mori model is adopted as the benchmark while the other two used for comparison purposes.
Figure~\ref{fig:supernovamodels} shows the time evolution of the luminosity and average neutrino energy predictions of the Mori, Nakazato and failed SN models. 
The dominant reaction (about 90\% of all events) of supernova neutrinos with water is inverse beta decay (IBD), in which a proton and an electron antineutrino react to produce a positron and a neutron (${\rm p}^+ + \bar{\nu}_{\rm e} \rightarrow {\rm n} + {\rm e}^+$).  
We therefore consider only IBD when determining the cut criteria in Section~\ref{sec:event_select}.

\subsection{Backgrounds}
For this search the most serious source of background comes from muon-induced spallation: the decay products of radioactive nuclei produced in the dissociation of oxygen nuclei following the passage of cosmic ray muons. 
A variety of spallation isotopes are formed in muon-oxygen interactions with different half-lives and daughter particle energies (c.f.~\cite{Li:2014sea,Li:2015lxa,Super-Kamiokande:2015xra}). 
The resulting isotopes from a single muon or from multiple muons may conspire to produce a cluster of events that appear in the search window described below.
Fortunately these events are highly correlated with the timing and path of their preceding muon and can be removed efficiently using existing algorithms, referred to as the ``spallation cut'' below. These are described in~\citep{Hosaka:2005um,Super-Kamiokande:2015xra}.
This cut rejects about 90\% of spallation events between 4 and 20 MeV, while rejecting only 20\% of simulated supernova events across all energies.

Other, less frequent, spurious event clusters may be produced by problems in the front-end electronics, discharges at the PMT dynode or data acquisition troubles that repeat over short intervals and become background to the analysis. Data known to include such problematic events are removed in the present search, although transient incidents falling in an otherwise normal data set are possible. 
The event selection process automatically removes most of those problematic events, but others must be removed manually at the end of the analysis by visual scanning.

\section{Event Selection}
\label{sec:event_select}

At the first stage of the analysis, events are required to have a reconstructed vertex within the 22.5~kton fiducial volume defined as the region of the ID more than 2~m from any of its walls.
At a later stage of the analysis, following the cluster search described below, we also use events outside of this volume to potentially augment the signal from a supernova candidate.
Events passing the initial fiducial volume cut are then subject to a test on their reconstructed quality.
The cut uses a goodness parameter based on the width of the PMT residual timing distribution to the reconstructed vertex, $g_t$ in~\citep{Cravens:2008aa}, and a second goodness parameter testing the azimuthal symmetry of the reconstructed hit pattern, $g_p$ in ~\citep{Abe:2016nxk}. 

The definition of $g_t$ combines two timing residuals 
for the vertex $\vec{v}$ as: 
\begin{equation}
    g_t =  \sum w_i e^{-\frac{1}{2}\left(\left(\frac{t_i-\tau_i(\vec{v})-t_0}{\omega}\right)^2+\left(\frac{t_i-\tau_i(\vec{v})-t_0}{\sigma}\right)^2\right)},
    \label{eq:g_t}
\end{equation}
with
\begin{equation}
    w_i = \frac{1}{ e^{-\frac{1}{2}\left(\frac{t_i-\tau_i(\vec{v})-t_0}{\omega}\right)^2}},
\end{equation}
where $\omega=60~{\rm ns}$ and $\sigma=5~{\rm ns}$.
Here $t_i$ is the time of the hit, $\tau_i(\vec{v})$ is the time-of-flight hit time assuming direct travel from the vertex to the $i^{th}$ PMT and $t_0$ is the light emission time. See \citep{Cravens:2008aa} for details.
SK observes Cherenkov light from charged particles as rings projected on its wall. A goodness parameter is used to study the uniformity of the PMT hit pattern around this ring.
This parameter, $g_p$, is defined as
\begin{equation}
   g_p = \frac{\max_i \left\{\angle_{\rm uni}(i)-\angle_{\rm data}(i)\right\}-\min_i \left\{\angle_{\rm uni}(i)-\angle_{\rm data}(i)\right\}}{2\pi},
   \label{eq:g_p}
\end{equation}
where $\angle_{\rm data}(i)$ is the azimuthal angle of the $i^{th}$ hit PMT included in a moving 50~ns time window
and $\angle_{\rm uni}(i)$ is the angle when a uniform PMT hit distribution is assumed. 
The origin of this angle is taken to be along the perpendicular line connecting the reconstructed direction with the earliest hit PMT in the ring.
The total fit quality is formulated as $g_t^{2} -g_p^{2}$ and ranges from 0 to 1, where 1 represents better fit quality. Simulated supernova neutrinos have an average fit quality of 0.45 whereas for spallation events the average fit quality is near 0.25. Based on this, events are required to have a fit quality greater than 0.25. See Appendix ~\ref{sec:perfomance_ovaq} for more detail.
Finally, since events in a supernova neutrino event cluster are expected to be separated from one another by $O(1~{\rm ms})$ and some spallation isotopes have very short lifetimes, the search requires a time difference of more than $50~\mu$s between all events. 
This cut predominantly removes most Michel electrons from untagged muons.

\begin{figure}[hbt]
    \centering
    \begin{tabular}{cc}
         \begin{minipage}{0.5\hsize}
         \centering
         \includegraphics[width=8cm]{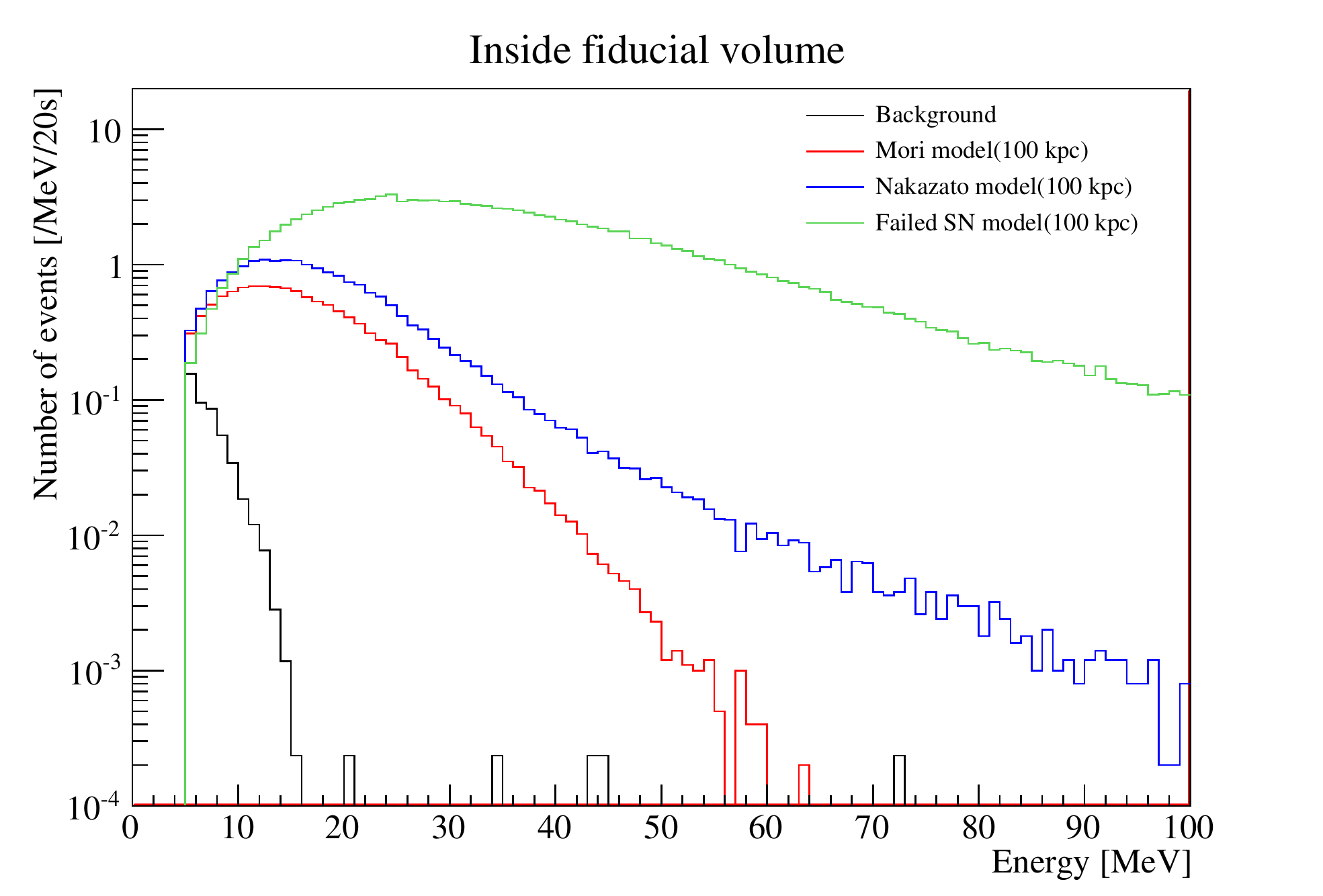}
         \end{minipage}&
         \begin{minipage}{0.5\hsize} 
         \centering
         \includegraphics[keepaspectratio, width=8cm]{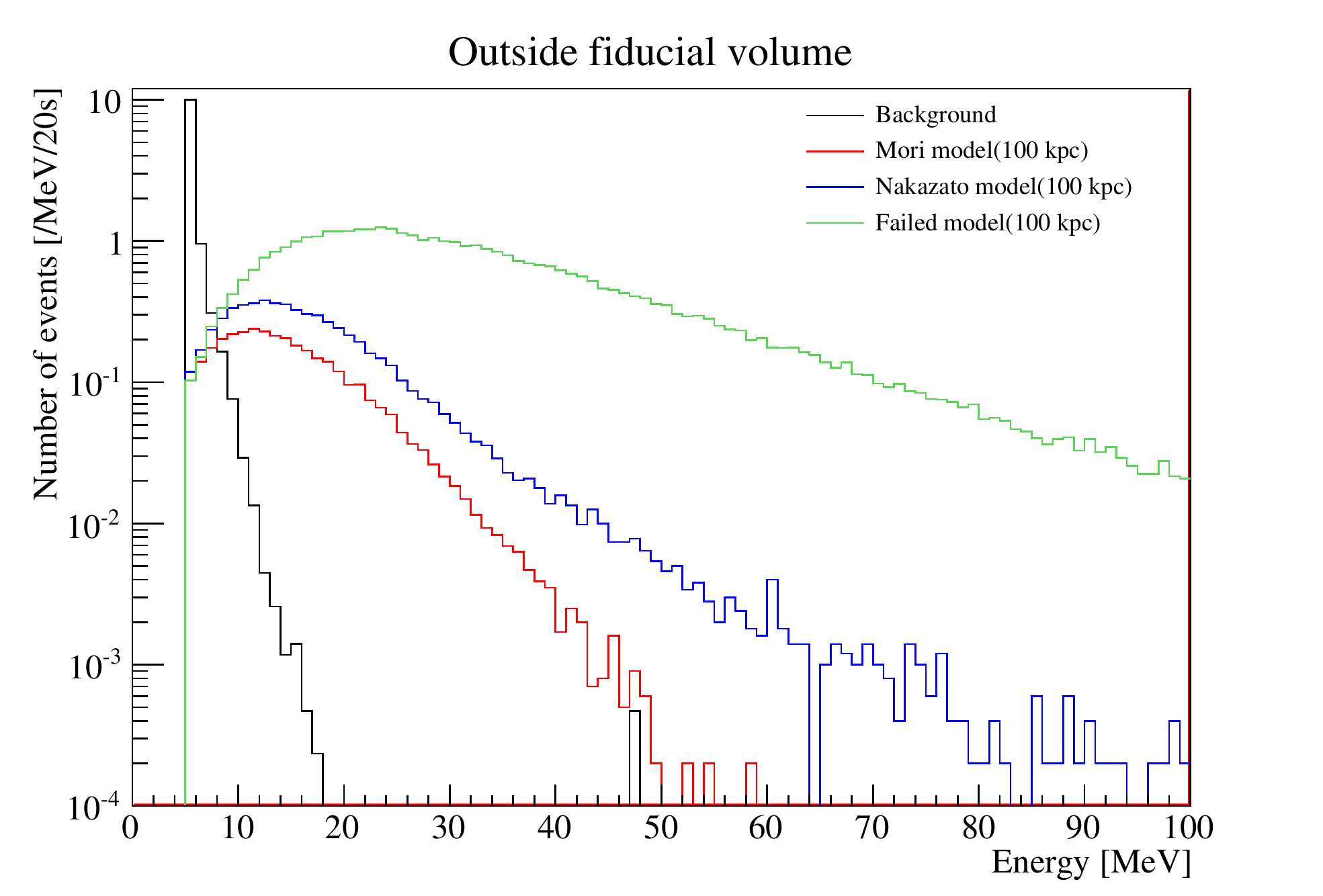}
         \end{minipage}
    \end{tabular}
    
    \caption{The spectra of background and supernovae MC at 100 kpc for the Mori and Nakazato models inside and outside the fiducial volume after the event selection in Table~\ref{tab:cut_criteria_in_FV}. Here data from a single 24 hour run from August 2011 is used for the background and we use only IBD interactions in the signal models (black lines). The solid lines show the result before the spallation cut and the dashed lines show the result after the spallation cut.
At present there is no spallation cut for the region outside the fiducial volume.
Red lines show the Mori model and blue lines show the Nakazato model.
All histograms have been normalized to the expected number of events in 20 s.\label{fig:energy_spectra}}
\end{figure}

Figure~\ref{fig:energy_spectra} shows simulated energy spectra 
from different models integrated over the first 20~s and an ``estimated background'', consisting of data from a single 24-hour run in 2011 that is assumed to contain no supernova events.
The figure shows events after selection inside (top left) and outside of the fiducial volume (top right).
These show that most of the background events accumulate at energies below 15 MeV. 
The bottom panel of the figure shows the events inside the fiducial volume after applying the spallation cut. 
Remaining events in the background can be attributed to spallation events that passed the selection as well as other types of events, such as solar neutrinos or electronics noise.
Note that the spallation cut has not been tuned for the region outside of the fiducial volume so will not be used there in this analysis. The figure shows the non-supernova event rate is less than $10^{-2}$ events over 20~s above 5.5~MeV inside the fiducial volume. Similarly, outside the fiducial volume and above 15~MeV the event rate is also less than $10^{-2}$. 
These energies are chosen as additional cuts applied to these two regions.
The cuts for this stage of the event selection are summarized in Table~\ref{tab:cut_criteria_in_FV}.

\begin{table}
    \centering
    \begin{tabular}{|c|cc|}
        \hline
         &Inside fiducial volume& Outside fiducial volume\\
         Vertex distance to ID wall & $> 200 {\rm\ [cm]}$ & $0 < d < 200 {\rm\ [cm]}$\\
         \hline
         Time difference to previous event & $> 50{\rm\ [\mu\sec]}$ & $> 50{\rm\ [\mu\sec]}$\\
         Energy  &  $> 5.5{\rm\ [MeV]}$ & $> 15{\rm\ [MeV]}$\\
         Fit quality  & $\rm >\ 0.25$ & $\rm > 0.25$\\
         \hline
    \end{tabular}
    \caption{Event selection criteria for the input to the cluster search. The fiducial volume is defined as the region 200~cm inside of the ID wall. Here the energy cut refers to the total energy of positron candidates. \label{tab:cut_criteria_in_FV}}
\end{table}

\subsection{Cluster Search}\label{sec:time_window}
Events surviving the pre-selection above are passed to a search algorithm that identifies clusters of events occurring within any of three specified time windows. The limits of these time windows are chosen based on the evolution of the neutrino flux during a supernova using criteria from the previous study~\citep{Ikeda:2007sa}. The 0.5~s window corresponds to the time between the initial collapse and subsequent bounce. The 2~s window covers the time until the shock is revived. The 10~s window corresponds to the neutron star cooling phase. In order to enhance the analysis efficiency, the event thresholds for these time windows are roughly half those used in the previous study, requiring at least 2, 2, and 4 events per cluster, respectively. The time windows are shown in Table~\ref{tab:time_window}. These criteria were analyzed to ensure they increased sensitivity without increasing the rate of background clusters.
The number of events forming a cluster identified by the selection criteria is referred to as the cluster's multiplicity.
See Section~\ref{sec:MC} for details.

\begin{table}[htb]
    \centering
    \begin{tabular}{cc}
         \hline
            Time window 1 & $\geq$ 2 events in ${\rm 0.5\ [s]}$ \\
            Time window 2 & $\geq$ 2 events in ${\rm 2\ [s]}$ \\
            Time window 3 & $\geq$ 4 events in ${\rm 10\ [s]}$ \\
         \hline
    \end{tabular}
    \caption{Time window settings.\label{tab:time_window}}
\end{table}

\subsection{Spatial Classification and Cluster Cuts}
\label{sec:ml_cluster}
For a nearby supernova, the spatial distribution of events in the detector is expected to be uniform within its volume. In contrast, spallation background events are expected to be distributed along their parent muon's track. When the event multiplicity in a cluster is low the observed event distribution may deviate from these ideals. 
To account for this, clusters are separated into four spatial classifications--- volume-like, plane-like, line-like, and point-like--- 
based on the vertex distribution of their events. The algorithm uses the eigenvalues of the correlation matrix of vertices, 
$\langle (x_{i} - \langle x_{i} \rangle^{2})(x_{j} - \langle x_{j} \rangle^{2})\rangle$, 
where $i$ and $j$ label axes of the vectices, to determine which of the geometries is most consistent with the vertex distribution. 
More details can be found in \citep{Abe:2016waf}.

Clusters are then subject to different cuts depending upon their spatial classification. 
Three variables are introduced for this purpose: 
the average kinetic energy of its reconstructed positron candidates, $\average{E_{\rm kin}}$, 
the average pairwise distance between event vertexes in a cluster, $\average{D}$, and the average residual distance of event vertexes to the cluster centroid, $\average{\Gamma}$.  
They are  defined as, 
\begin{equation}
\average{E_{\rm kin}} = \frac{\sum^{M}_{i=1}E_{i}}{M},
\end{equation}

\begin{equation}
    \average{D} = \frac{\sum^{M}_{i<j} |\vec{d_i}-\vec{d_j}|}{_{M}C_{2}},
\end{equation}

\begin{equation}
        \average{\Gamma} = \sqrt{\frac{1}{3\left(M-1\right)}\sum^{M}_{i=1}\left(\vec{d_i} - \vec{d_0}\right)^2},
\end{equation}
where $M$ is the number of events included in a cluster (the  multiplicity), $E_{i}$ is the total positron candidate energy of the $i$th event
in a cluster, $\vec{d_i}$ is the vertex of that event, $_{M}C_2$ is the number of combinations of $M$ taken 2 at a time, and 
$\vec{d_0} = 1/M\times\sum^{M}_{i=1}\vec{d_i}$ is the centroid of the cluster.

Clusters from spallation events are expected to have lower 
$\average{E_{\rm kin}}$ and smaller $\average{D}$ than supernova clusters. 
Spallation clusters typically have $\average{D}$ less than 1000 cm and
 $\average{E_{\rm kin}}$ less than 10 MeV while supernova clusters have larger values.
However, spallation clusters can have 
$\average{D}$ larger than 2000 cm for muon tracks traversing the entire length of the Super-K ID and producing multiple spallation isotopes.
Similarly, a group of muons from a single cosmic ray shower  bundles can produce several spallation events across the detector resulting in large values of $\average{D}$.

Clusters are divided into six types, based on their spatial classification and the number of events they contain, as  summarized in Table~\ref{tab:cluster_cut}. 
Low multiplicity line-like and point-like clusters were found to have sufficiently different distributions 
in the variables above to warrant separate cuts from their higher multiplicity counterparts. 
In particular, for point-like clusters with only 3 events, signal and background can be efficiently separated without using $\average{E_{\rm kin}}$, thereby reducing the dependence of cut tuning on the choice of supernova model.

\begin{table}[H]
    \centering
    \scalebox{0.70}[0.9]{
    \begin{tabular}{ccccccc}
    \hline \hline
         \multirow{2}{*}{Category} &  Volume-like & Plane-like & Line-like & Line-like & Point-like & Point-like \\
          & & &  Low Multiplicity & High Multiplicity  & Low Multiplicity & High Multiplicity \\ 
         Cut variables& $\average{D}$ vs $\average{E_{\rm kin}}$ & $\average{D}$ vs $\average{E_{\rm kin}}$ & $\average{D}$ vs $\average{E_{\rm kin}}$ & $\average{D}$ vs $\average{E_{\rm kin}}$ & $\average{D}$ vs $\average{E_{\rm kin}}$ & $\average{D}$ vs $\average{\Gamma}$ \\
         \hline
         
         $M$ & - & - &  $\leq 3$ & $\geq 4$ & $\leq 3$ &  $\geq 4$ \\
         $\average{D}$ & - & - & - & - & $\geq 500\,{\rm cm}$ for $M \geq 3 $  & or $\leq 500\,{\rm cm} $ for $M\leq 3$ \\
         \hline

          \hline
    \end{tabular}}
    \caption{The six cluster categories and their additional cut criteria.}
    \label{tab:cluster_cut}
    
\end{table}

Cuts on the variables in each of the six types have been optimized using machine learning (ML) methods provided by the ``Scikit-learn'' Python package~\citep{JMLR:v12:pedregosa11a}. The logistic regression (LR) and support vector machine (SVM) methods were used because they provide the highest accuracy (score) for correctly identifying signal and background clusters~\citep{10.5555/1162264}.
When performing the optimization, a sample of clusters from 200 days of data (tagged as spallation-like by the spallation likelihood) serves as the background model.
The signal simulation uses 30,000 supernova clusters generated using the Mori model assuming distances between around 100~kpc and 500\,kpc and  
including neutrino oscillation effects for the normal mass hierarchy calculated according to~\citep{Dighe:1999bi}.
These signal clusters typically have multiplicities of between 3 and 7 events. 
During training the background and supernova samples were balanced for each cluster category with background events having a weight of 20 compared to a weight of 1 for signal events.

In order to account for systematic errors in the reconstruction during cut optimization, the number of signal 
and background clusters is artificially increased by a factor of ten. 
Each cluster was used to generate 9 additional clusters, drawn from a random Gaussian distributions centered at the parent cluster's $\average{E_{\rm kin}}$ and $\average{D}$ with widths corresponding to 
the expected resolution in these variables, $1.5{\rm\ MeV}$ and $30 {\rm\ cm}$, respectively.
Training samples for the optimization were formed from 80\% of the resulting distributions, with 20\% used for validating the optimized cuts (called the ``test'' sample).

\begin{figure}[h]
    \centering
    \begin{tabular}{cc}
         \begin{minipage}{0.5\hsize}
         \centering
         \includegraphics[width=8cm]{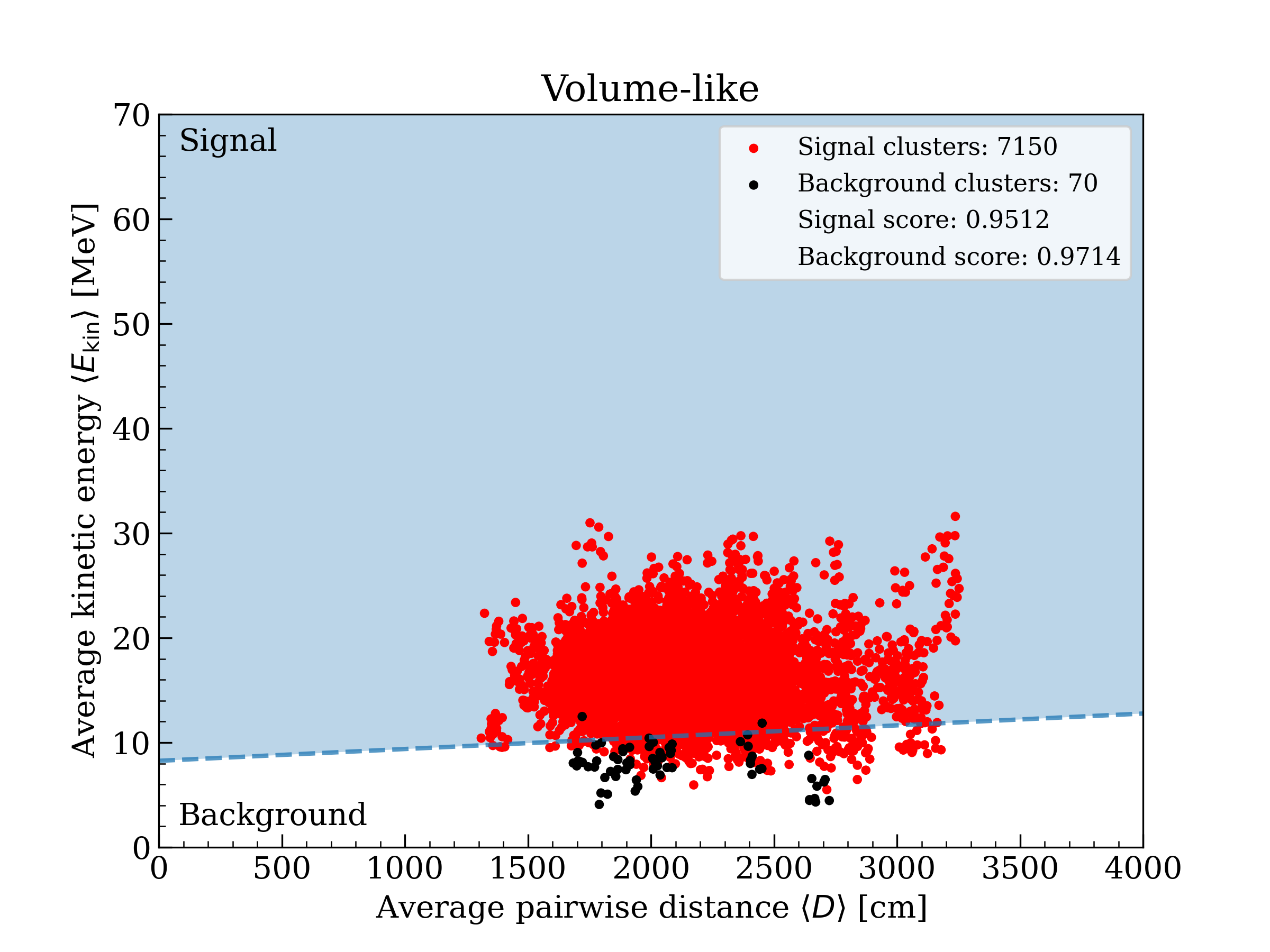}
         \end{minipage}&
         \begin{minipage}{0.5\hsize} 
         \centering
         \includegraphics[width=8cm]{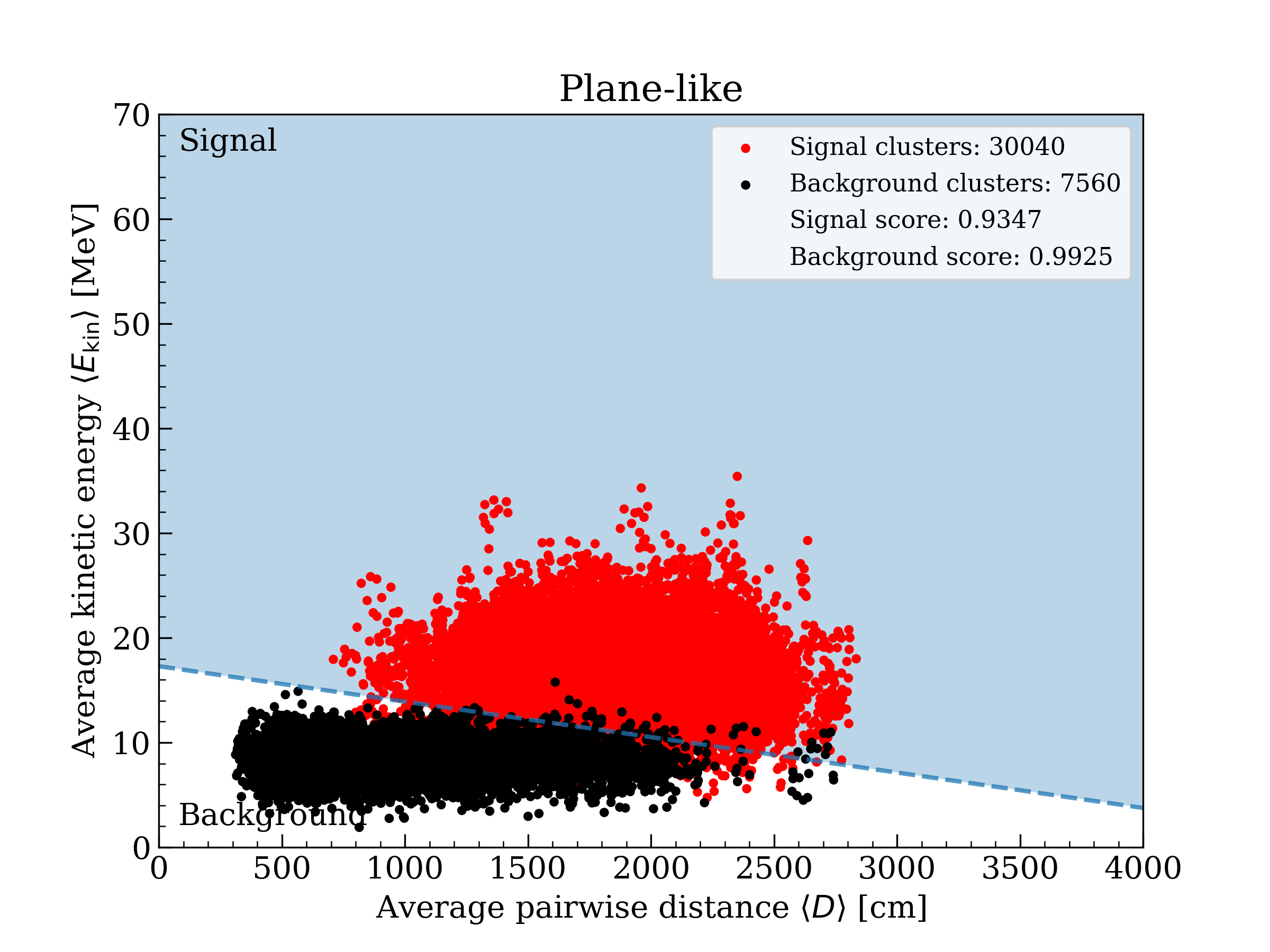}
         \end{minipage} \\
         
        \begin{minipage}{0.5\hsize}
         \centering
        \includegraphics[width=8cm]{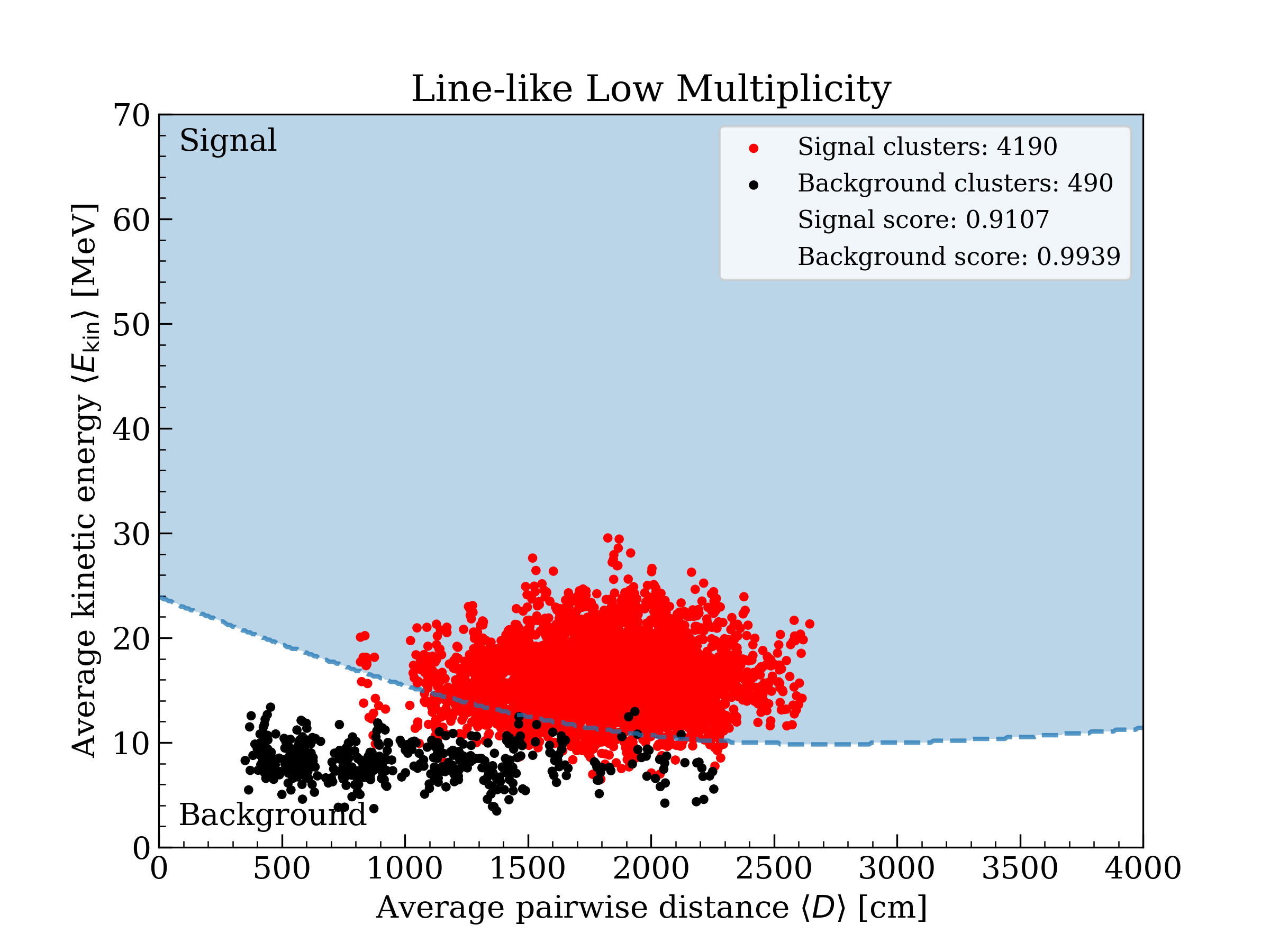}
         \end{minipage}&
         \begin{minipage}{0.5\hsize}
         \centering
         \includegraphics[width=8cm]{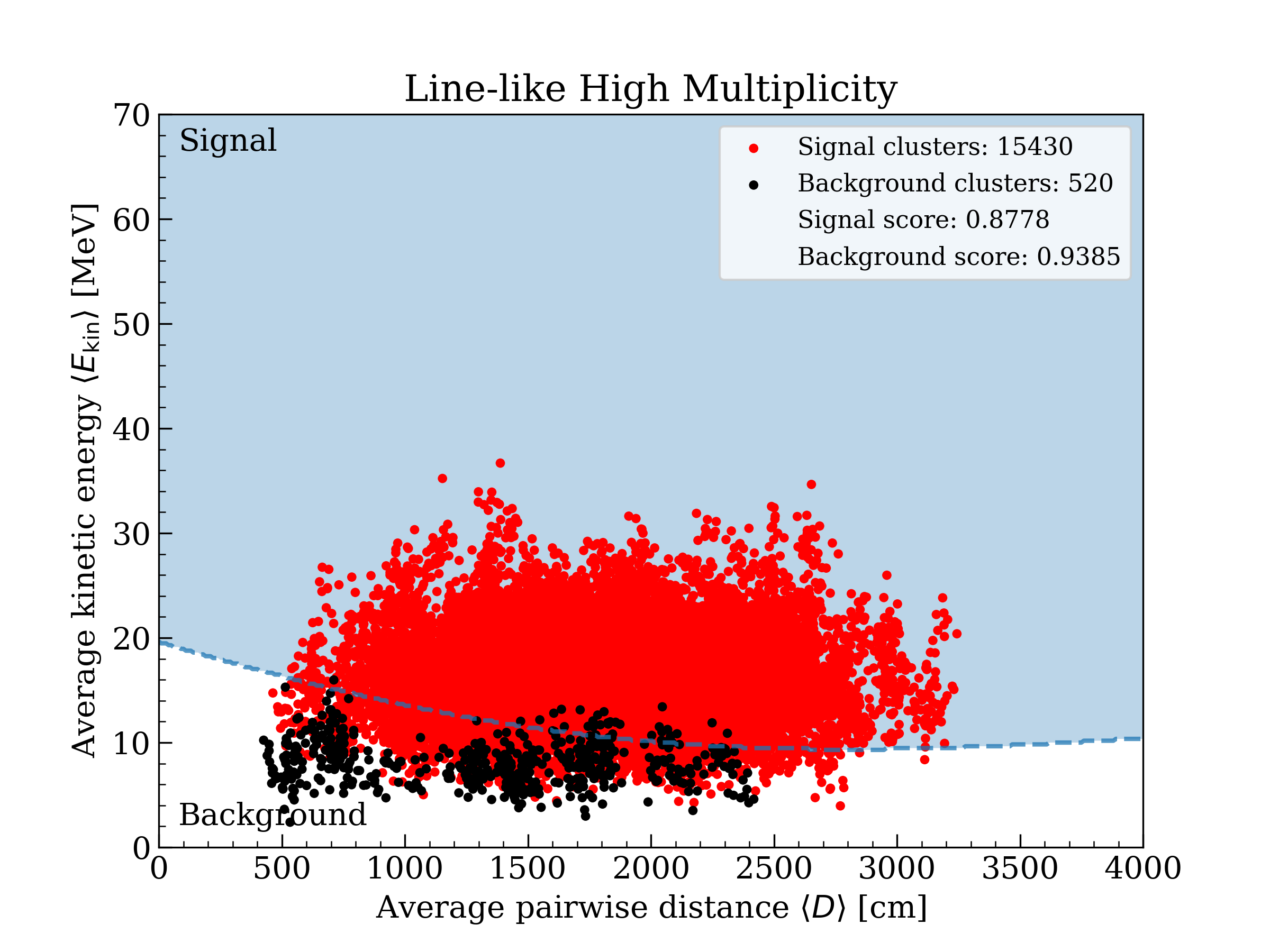}
         \end{minipage} \\
         \begin{minipage}{0.5\hsize}
         \centering
        \includegraphics[width=8cm]{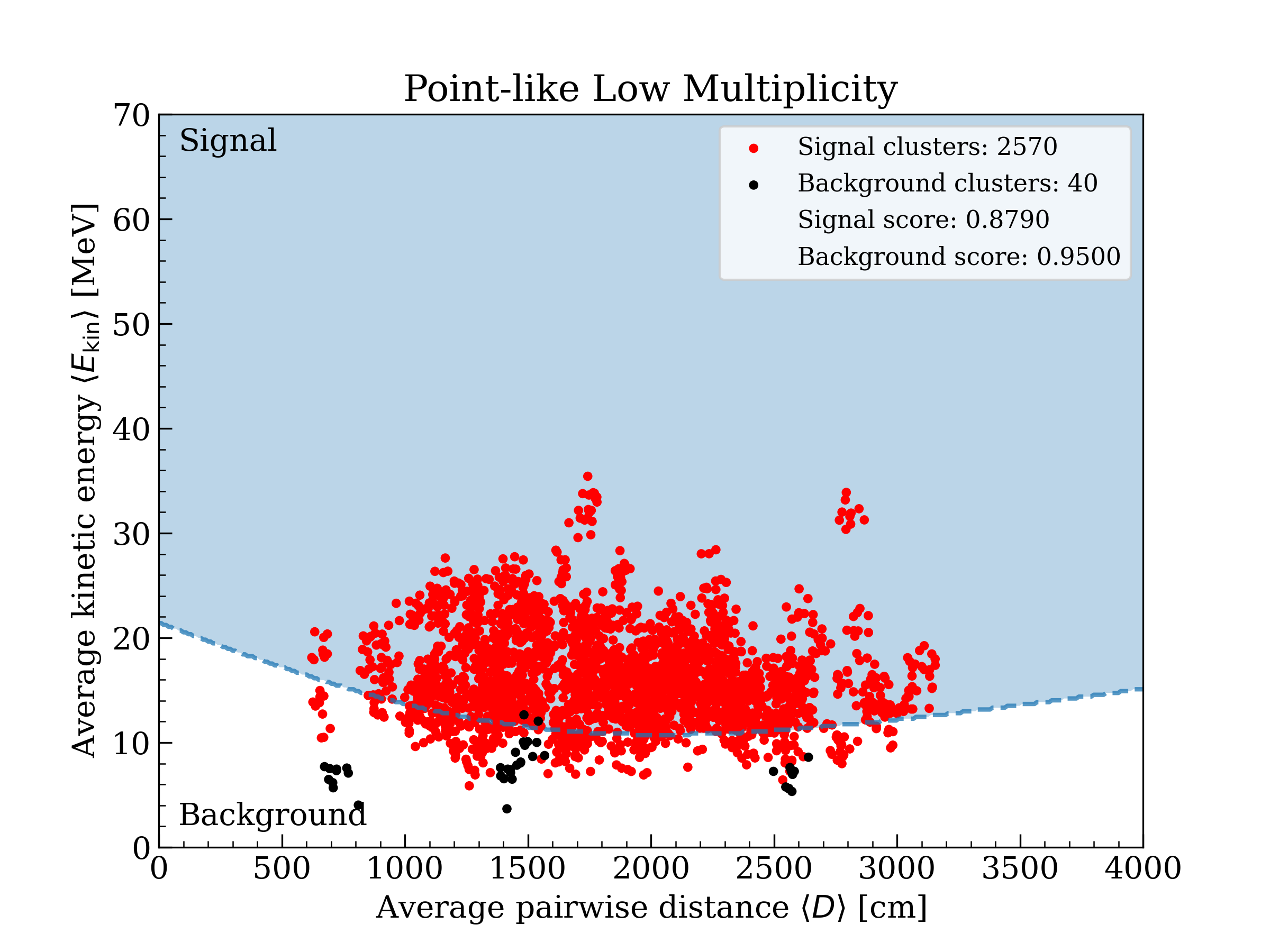}
         \end{minipage}&
         \begin{minipage}{0.5\hsize}
         \centering
         \includegraphics[width=8cm]{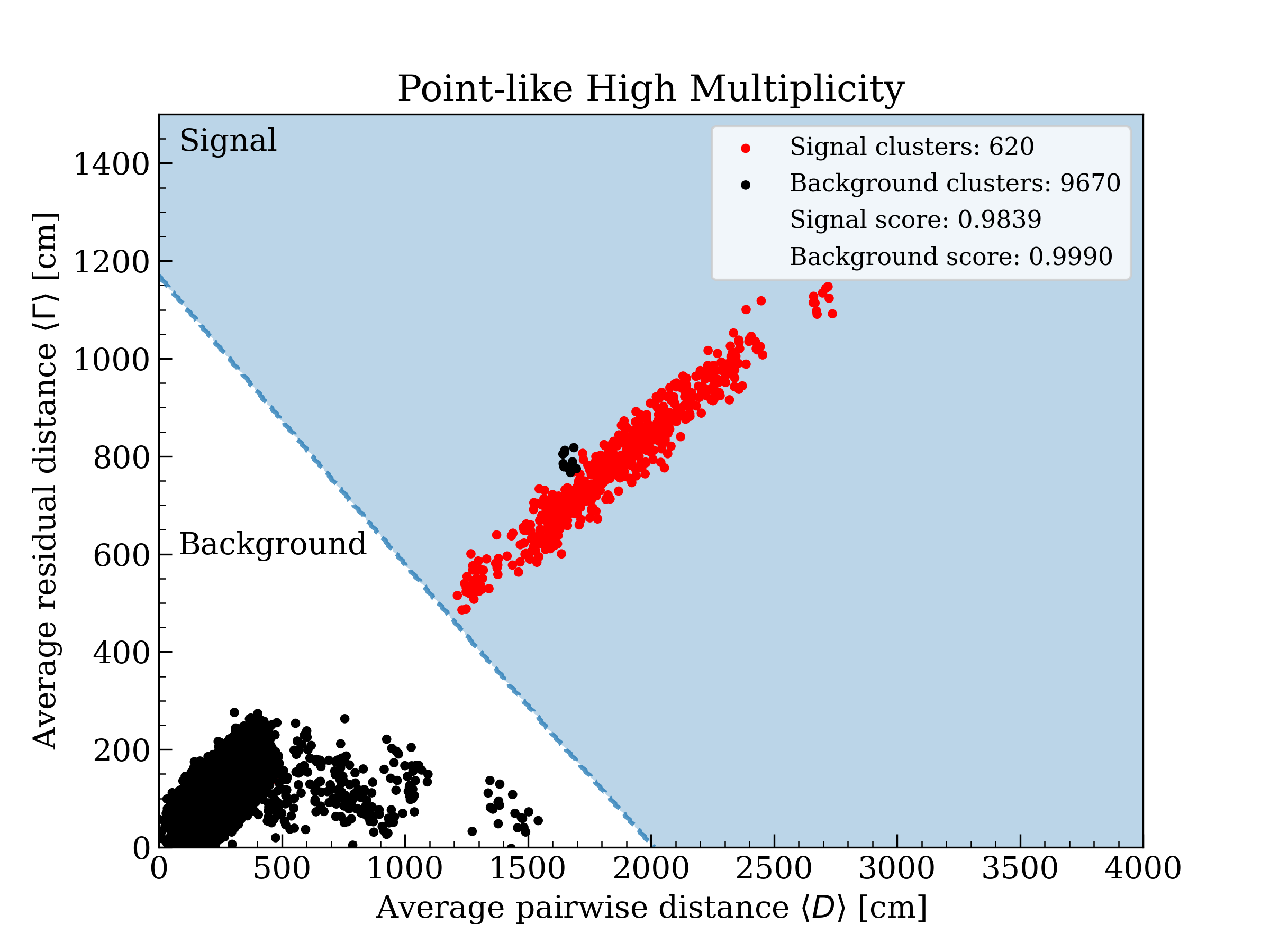}
         \end{minipage} \\
    \end{tabular}
    \caption{Optimized ML classification curves for the six cluster types and classification result for the test sample taken from 204 days of data. 
Red points denote the supernova signal and black points are background taken from 204 days of data. 
The blue regions are classified as signal-like while the white regions are background-like. Note that the spallation cut has not been applied. }
\label{fig:ml_train}
\end{figure}

Table~\ref{tab:ml_type} shows the background and signal classification scores after cut optimization for the six cluster types. The score represents the fraction of events of a given category, signal or background, that were correctly classified.  
When selecting a model for each category, preference was given to that with the highest background score. 
If multiple models had the same background score then the one with the highest signal score was chosen.  
Using this method the LR was adopted for the volume-like and plane-like categories and SVM for the others.
The ``weighted average'' column of the table represents the sum of scores across the signal (background) row weighted by the relative fraction 
of signal (background) clusters in each column of that row. 
These weights are shown in the rows labeled as ``Fraction''.
The corresponding signal and background scores are 0.994 and 0.919, respectively.
However, we note that this information is simply added for reference and was not used in the selection of ML models.
Figure~\ref{fig:ml_train} shows the results of the cut optimization overlaid 
with the test signal and background samples. The black cluster in the signal region in the bottom right panel of Figure~\ref{fig:ml_train} is due to   spallation events produced by multiple muons. As seen in the figure such point-like clusters are rare and are easily removed with the spallation cut (c.f. Figure~\ref{fig:results}).

\begin{table}[H]
    \centering
    \scalebox{0.70}[0.9]{
    \begin{tabular}{cccccccc}
    \hline \hline
        \multirow{2}{*}{Category} &  Volume-like & Plane-like & Line-like & Line-like & Point-like & Point-like &   \multirow{2}{*}{Weighted Average} \\ 
                           &   &        &  Low Multiplicity & High Multiplicity  & Low Multiplicity & High Multiplicity \\ 
         \hline
         ML model     & LR     & LR     & SVM    & SVM    & SVM    & SVM    & - \\
         
         Signal score & 0.951 & 0.935 & 0.911 & 0.878 & 0.879 & 0.984 & 0.919 \\
         Fraction     & 0.119 & 0.501 & 0.070 & 0.257 & 0.043 & 0.010 & 1.0     \\ 
         \hline
         Background score & 0.971 & 0.993 & 0.994 & 0.939 & 0.950 & 0.999 & 0.994 \\
         Fraction         & 0.004 & 0.412 & 0.027 & 0.028 & 0.002 & 0.527 & 1.0     \\ 
         \hline
         \hline
    \end{tabular}}
    \caption{Machine learning models for each cluster category and resulting scores. Fraction denotes the fraction of the corresponding sample in each category. The Weighted Average column is the sum of scores weighted by those fractions. \label{tab:ml_type}}
\end{table}

Clusters that are classified as signal-like by the ML method for their type are then passed to the spallation cut. \footnote{Since this cut is computationally expensive it is only applied to passing signal-like clusters.}
The spallation cut is applied on an event-by-event basis meaning that some clusters may contain both 
spallation-like events and other events. 
In these cases the spallation-like events are removed from the cluster, the input variables for the ML cut are recomputed, and the ML cut is reapplied.  
Any signal-like clusters remaining are then subject to further cross checks to reduce backgrounds from transient instabilities in the front-end electronics.
The search is then expanded to the outer fiducial volume around the time of any clusters remaining 
to determine if there are other events that should be associated with that cluster.

\section{Analysis performance}\label{sec:performance}

The effect of the event-by-event selection (discussed in Section~\ref{sec:event_select}) on the ``background'' from one day of the Super-K data and 
simulated supernovae for the Mori and Nakazato models are shown in Table~\ref{tab:reduction_summary}. A supernova distance of 1~kpc was assumed to give large data sets. The numbers in brackets show the  efficiency relative to the previous cut. 
The total efficiency, including all cuts, is 72.9\% (27.5\%) for the Mori model and 75.1\% (34.4\%) for the Nakazato model and background rejection is 99.3\% (99.9\%) inside (outside) the fiducial volume.

\begin{table}[htbp]
    \centering
    \begin{tabular}{ccc}
     \hline\hline
    \multicolumn{3}{c}{Background data}\\
    \hline
     & Inside fiducial volume & Outside fiducial volume\\
    \hline
    Reconstructed events & 40215 & 166680 \\
    Energy cut & 16018 (39.8\%) & 12 (0.01\%) \\
    Time difference cut & 40192 (99.9\%) & 166585 (99.9\%) \\
    Fitting quality cut & 2005 (5.0\%) & 49178 (29.5\%) \\
    Spallation cut & 17621 (43.8\%) &  - \\
    All cuts & 297 (0.74\%) &  12(0.01\%) \\
    \hline\hline
    \multicolumn{3}{c}{Mori model at 1~kpc}\\
    \hline
     & Inside fiducial volume & Outside fiducial volume\\
    \hline
    True events & 126067 & 55987 \\
    Reconstructed events & 123995 (98.4\%) & 39044 (69.7\%) \\
    Energy cut & 116287 (92.2\%) & 15505 (27.7\%) \\
    Time difference cut & - & - \\
    Fitting quality cut & 119521 (94.8\%) & 37336 (66.7\%) \\
    Spallation cut & 99196 (78.7\%) &  - \\
    All cuts & 91864 (72.9\%) &  15396 (27.5\%) \\
    \hline\hline
    \multicolumn{3}{c}{Nakazato model at 1~kpc}\\
    \hline
     & Inside fiducial volume & Outside fiducial volume\\
    \hline
    True events & 177847 & 78738 \\
    Reconstructed events & 175197 (98.5\%) & 55396 (70.4\%) \\
    Energy cut & 168769 (94.9\%) & 27293 (34.7\%) \\
    Time difference cut & - & - \\
    Fitting quality cut & 170772 (96.0\%) & 53520 (62.5\%) \\
    Spallation cut & 140157 (78.8\%) &  - \\
    All cuts & 133553 (75.1\%) &  27113 (34.4\%) \\
    \hline\hline
    \end{tabular}
    \caption{Reduction summary using one day of data and simulated SN events at 1~kpc using the Mori and Nakazato models. 
Cut criteria are shown in Table. \ref{tab:cut_criteria_in_FV}.}
\label{tab:reduction_summary}
    
\end{table}

\subsection{Monte Carlo simulation of observation}\label{sec:MC}

\begin{figure}[H]
    \centering
    \begin{tabular}{cc}
         \begin{minipage}{0.5\hsize}
         \centering
         \includegraphics[width=8cm]{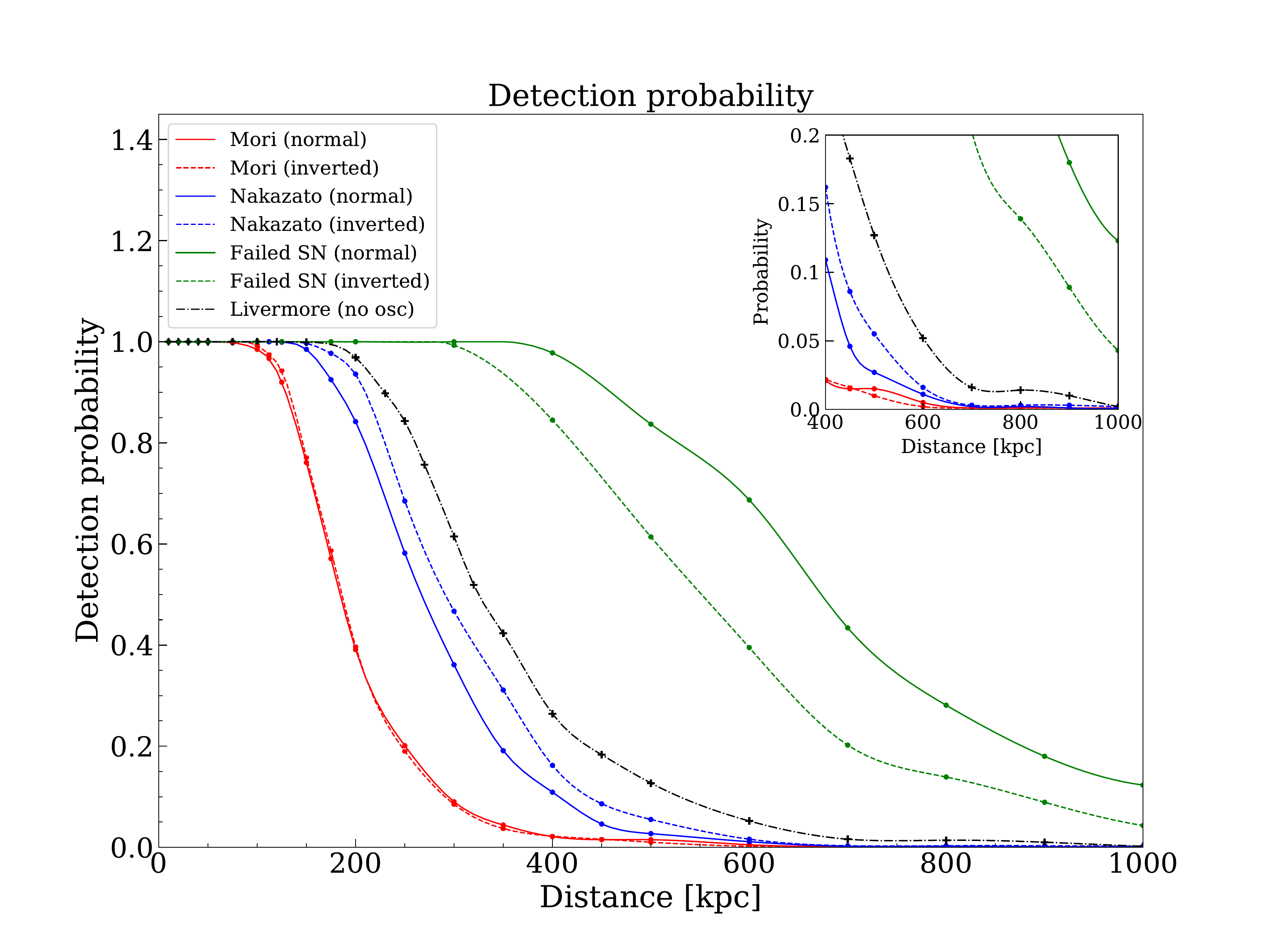}
         \end{minipage}&
         \begin{minipage}{0.5\hsize} 
         \centering
         \includegraphics[keepaspectratio, width=8cm]{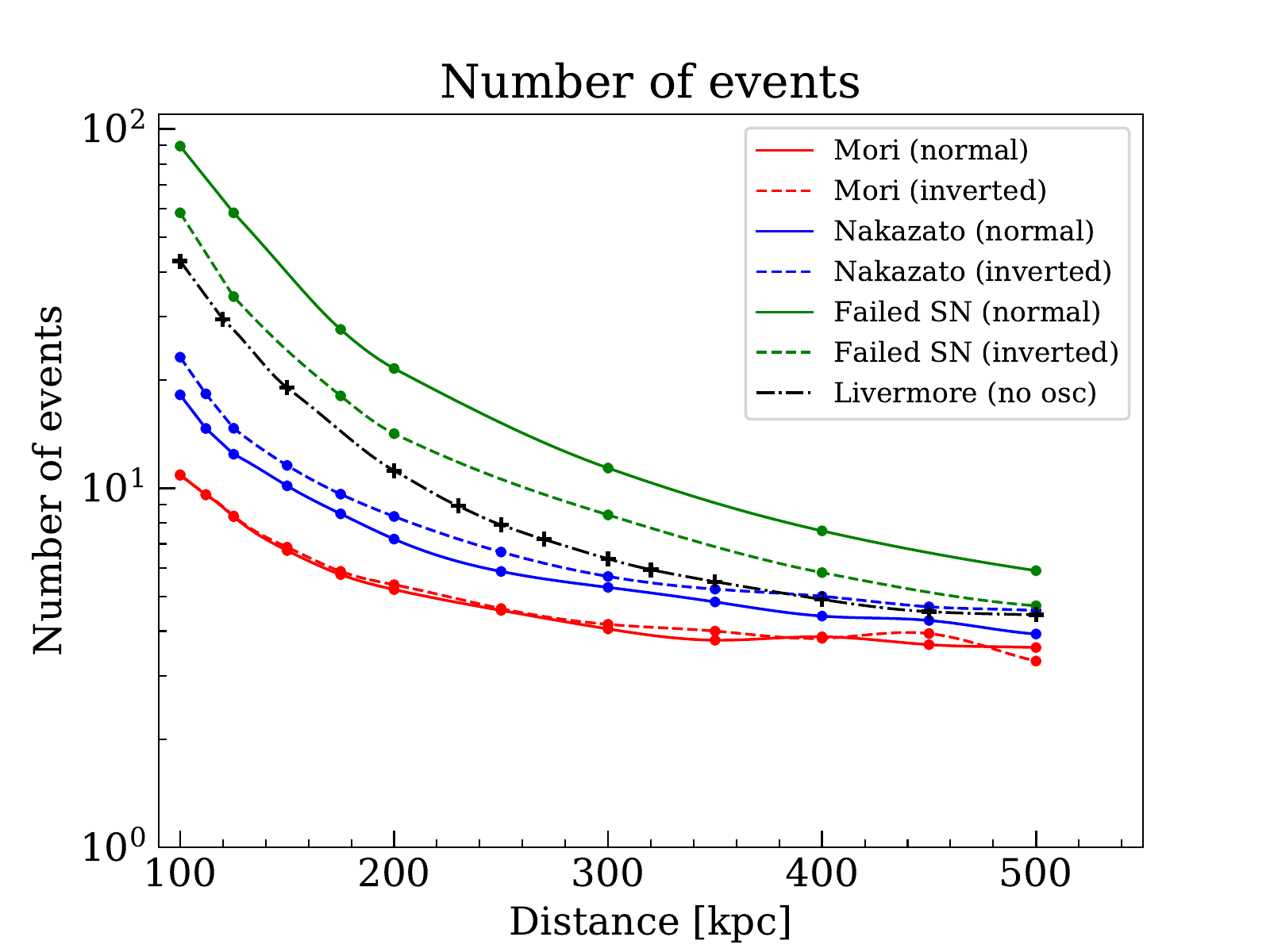}
         \end{minipage}
    \end{tabular}
    
    \caption{Predicted detection probability and number of detected events as a function of distance for several models.}
\label{fig:mc_predict}
\end{figure}

In order to estimate the expected event rate and analysis efficiency as a function of the supernova distance, several 
simulated supernovae were generated from the Mori and Nakazato models assuming oscillations under both the 
normal and inverted mass hierarchies~\citep{Dighe:1999bi}. 
\footnote{Note that this model of oscillations assumes adiabatic flavor conversions, though other transitions are also possible.}
The failed supernova model from Nakazato, in which the star collapses directly to a black hole, 
and the Livermore model were also simulated. 
Here 1000 supernovae were simulated at each assumed distance. 
Figure~\ref{fig:mc_predict} shows the detection probability $P_{\rm detect}$, 
\begin{equation}
    P_{\rm detect} = \frac{N_{\rm detected}}{N_{\rm sim}},
\end{equation}
\noindent where $N_{\rm detected}$ is the number of detected simulations, $N_{\rm sim}$ is the total number of simulations.
Here``detected'' means a simulation that includes at least one event cluster satisfying the cluster criteria described in Section~\ref{sec:ml_cluster}. 

The left side of Fig. \ref{fig:mc_predict} shows that  
the detection probability for all models is 1.0 up to at least 100~kpc. 
Further, in the present study the non-zero detection probability extends to larger distances than that of the previous analysis, which was based on the Livermore model, with a similar energy threshold.
For the Mori and Nakazato models the detection probability drops to a few percent at around 500~kpc, representing the limit of the present search's sensitivity.
Detection probabilities for all models are also presented in Table~\ref{tab:detection_probability}.

\begin{table}[H]
    \centering
    \scalebox{0.75}[0.9]{
    \begin{tabular}{ccccccccccc}
    \hline \hline
         Model &  100\,kpc & 200\,kpc & 300\,kpc & 400\,kpc & 500\,kpc & 600\,kpc & 700\,kpc & 800\,kpc & 900 kpc & 1000\,kpc  \\
         \hline
        Mori (Normal)& 0.985& 0.391& 0.090& 0.021& 0.015& 0.005& 0.001& 0.001& 0.001& 0.001\\
        Mori (Inverted)& 0.993& 0.397& 0.085& 0.022& 0.010& 0.002& 0.001& 0.001& 0.001& 0.001\\
        Nakazato (Normal)& 1.000& 0.842& 0.361& 0.109& 0.027& 0.011& 0.002& 0.002& 0.001& 0.001\\
        Nakazato (Inverted)& 1.000& 0.936& 0.467& 0.162& 0.055& 0.016& 0.003& 0.003& 0.003& 0.002\\
        Failed SN (Normal)& 1.000& 1.000& 1.000& 0.978& 0.837& 0.687& 0.434& 0.281& 0.180& 0.123\\
        Failed SN (Inverted)& 1.000& 1.000& 0.993& 0.845& 0.614& 0.395& 0.202& 0.139& 0.089& 0.043\\
        Livermore (No osc)& 1.000& 0.969& 0.615& 0.264& 0.127& 0.052& 0.016& 0.014& 0.010& 0.002\\
         \hline

\hline
         \hline
    \end{tabular}}
    \caption{Detection probability for each model\label{tab:detection_probability} as a function of distance for several models and oscillation assumptions. These numbers are used in the left panel of Figure~\ref{fig:mc_predict}.}
\end{table}

The right side of Figure~\ref{fig:mc_predict} shows the total number of events summed over all selected clusters for supernovae detected with this method. 
On average 10 events are expected for the Mori model at 100~kpc and 15 events for the Nakazato model. This average does not follow an inverse square law with distance since at least two events are required 
to define a cluster (c.f. Table~\ref{tab:time_window}).
The predicted number of events for each model is summarized in Table~\ref{tab:prediction_events}. 

\begin{table}[H]
    \centering
    \begin{tabular}{cccccc}
    \hline\hline
         Model &  100 kpc & 200 kpc & 300 kpc & 400 kpc & 500 kpc\\
         \hline
            Mori (Normal)& 10.9& 5.2& 4.1& 3.9& 3.6\\
            Mori (Inverted)& 10.9& 5.4& 4.2& 3.8& 3.3\\
            Nakazato (Normal)& 18.2& 7.2& 5.3& 4.4& 3.9\\
            Nakazato (Inverted)& 23.2& 8.3& 5.7& 5.0& 4.6\\
            Failed SN (Normal)& 89.7& 21.5& 11.4& 7.6& 5.9\\
            Failed SN (Inverted)& 58.5& 14.2& 8.4& 5.8& 4.7\\
            Livermore (No osc)& 42.9& 11.2& 6.4& 4.9& 4.4\\
         \hline\hline
    \end{tabular}
    \caption{Prediction of the number of  events\label{tab:prediction_events} as a function of distance for several models and oscillation assumptions. These numbers are also used in the right panel of Figure~\ref{fig:mc_predict}.}
\end{table}

\section{Data analysis and Result}\label{sec:result}

All Super-Kamiokande data from the SK-IV period, representing 3318.41 days of livetime were used in the analysis. 
There were 151923 clusters found by the cluster search, 362 of which were classified as volume-like and
71 of which were plane-like.
Further, 4857 and 4753 clusters were classified as line-like Low Multiplicity and High Multiplicity, with 186 point-like Low Multiplicity and 80787 High Multiplicity clusters as shown in Table~\ref{tab:cluster_1st}.

\begin{figure}
    \centering
\begin{tabular}{cc}
         \begin{minipage}{0.5\hsize}
         \centering
         \includegraphics[width=8cm]{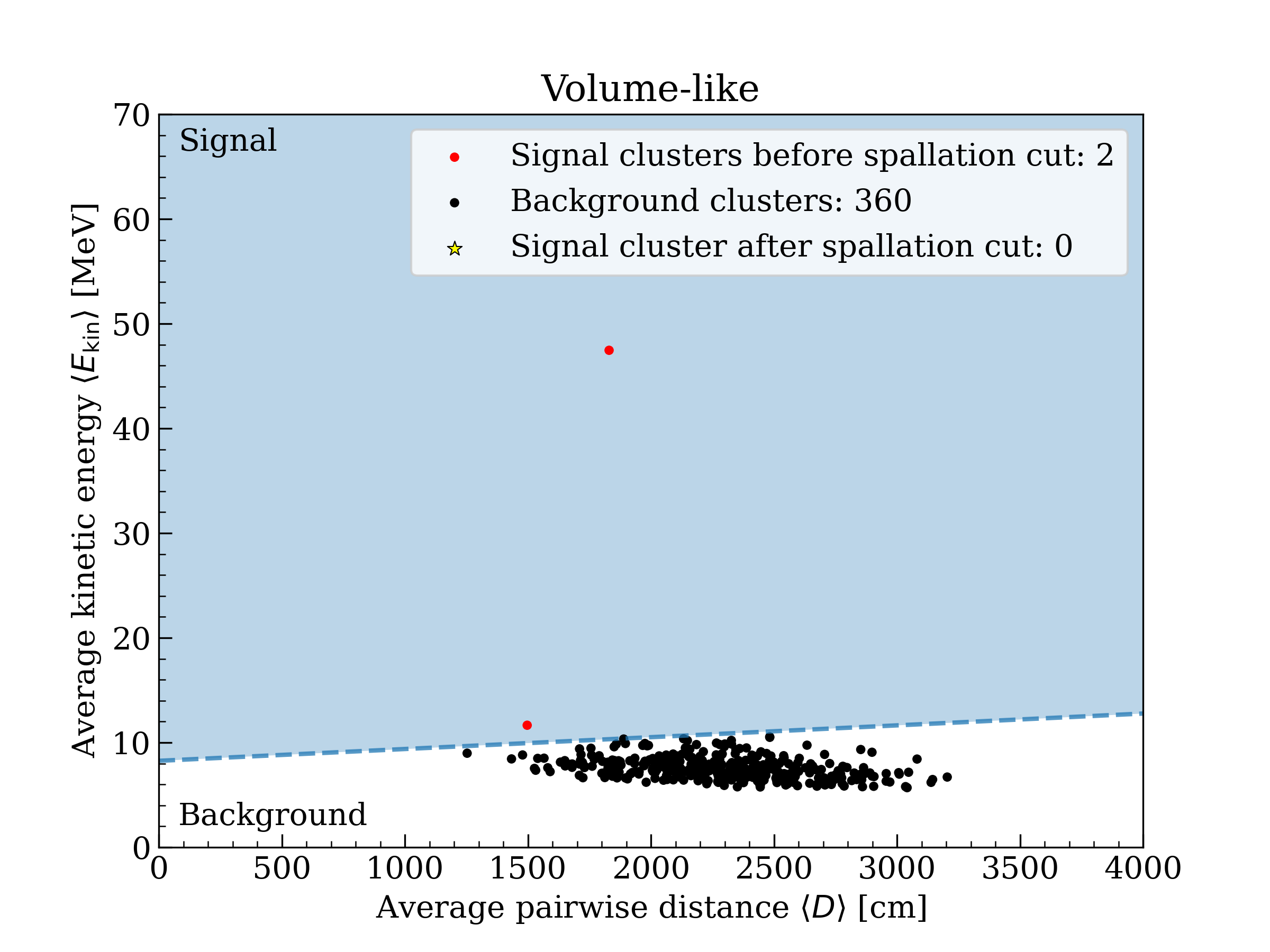}
         \end{minipage}&
         \begin{minipage}{0.5\hsize} 
         \centering
         \includegraphics[width=8cm]{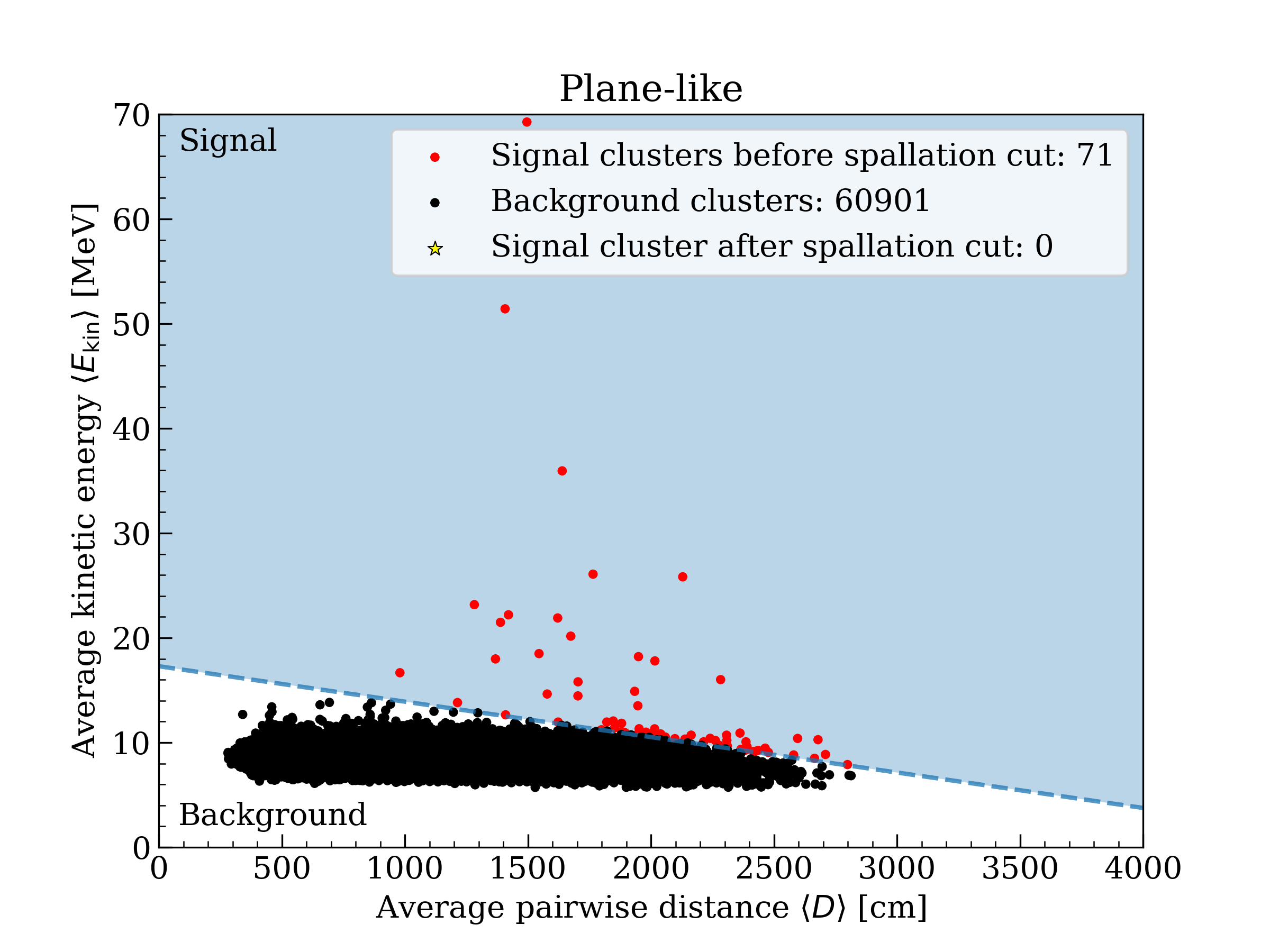}
         \end{minipage} \\
         
        \begin{minipage}{0.5\hsize}
         \centering
        \includegraphics[width=8cm]{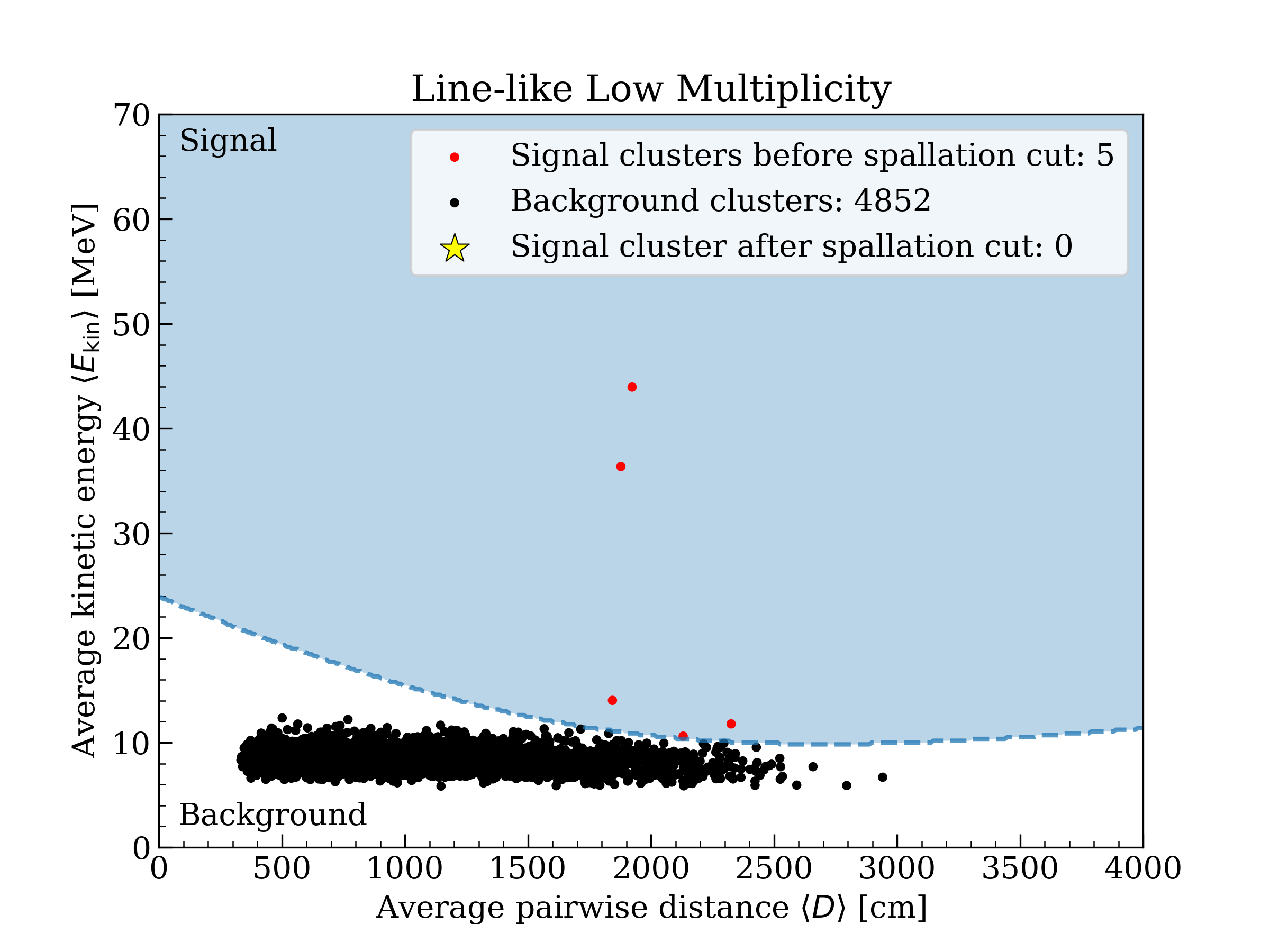}
         \end{minipage}&
         \begin{minipage}{0.5\hsize}
         \centering
         \includegraphics[width=8cm]{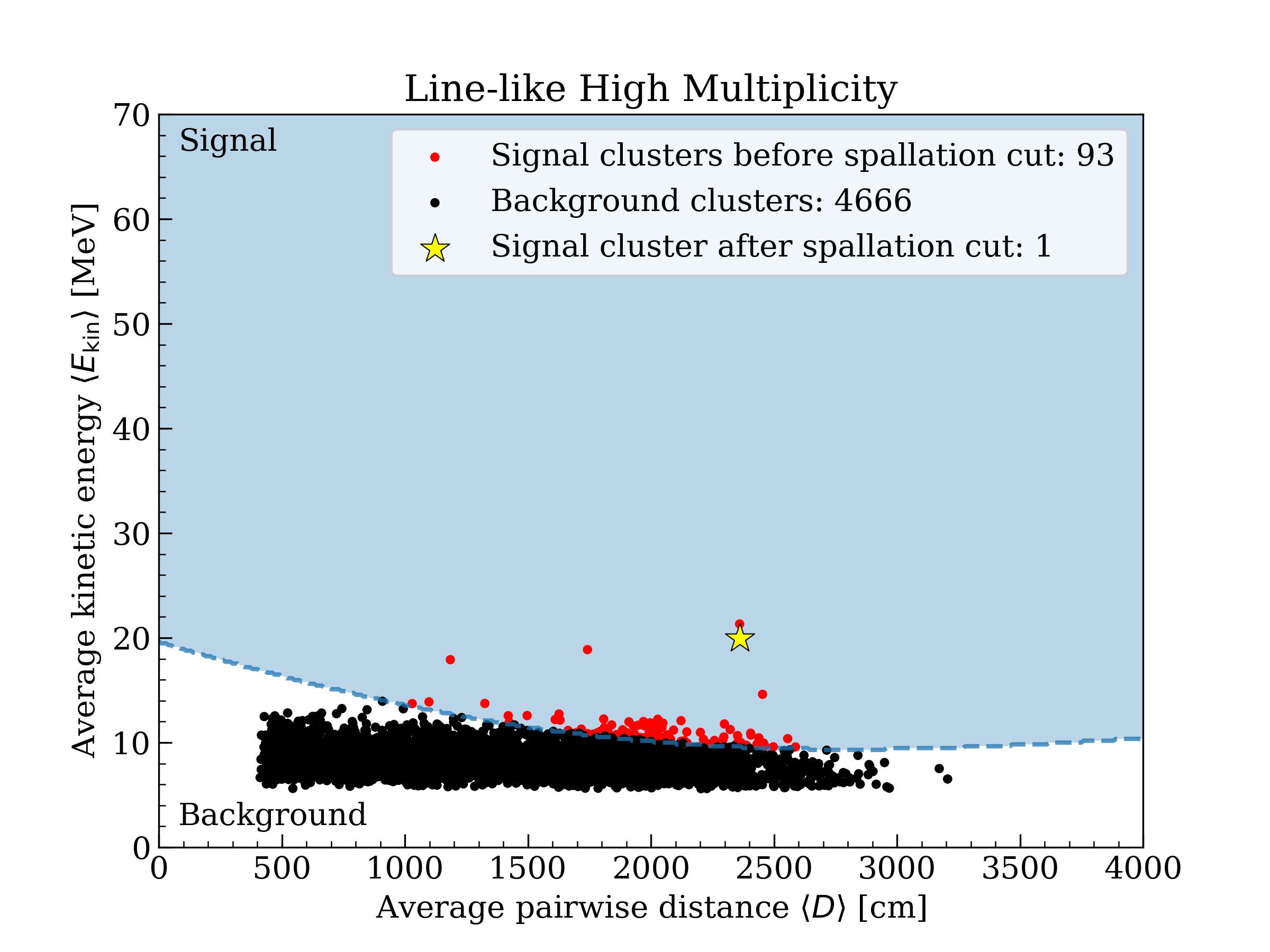}
         \end{minipage} \\
         \begin{minipage}{0.5\hsize}
         \centering
        \includegraphics[width=8cm]{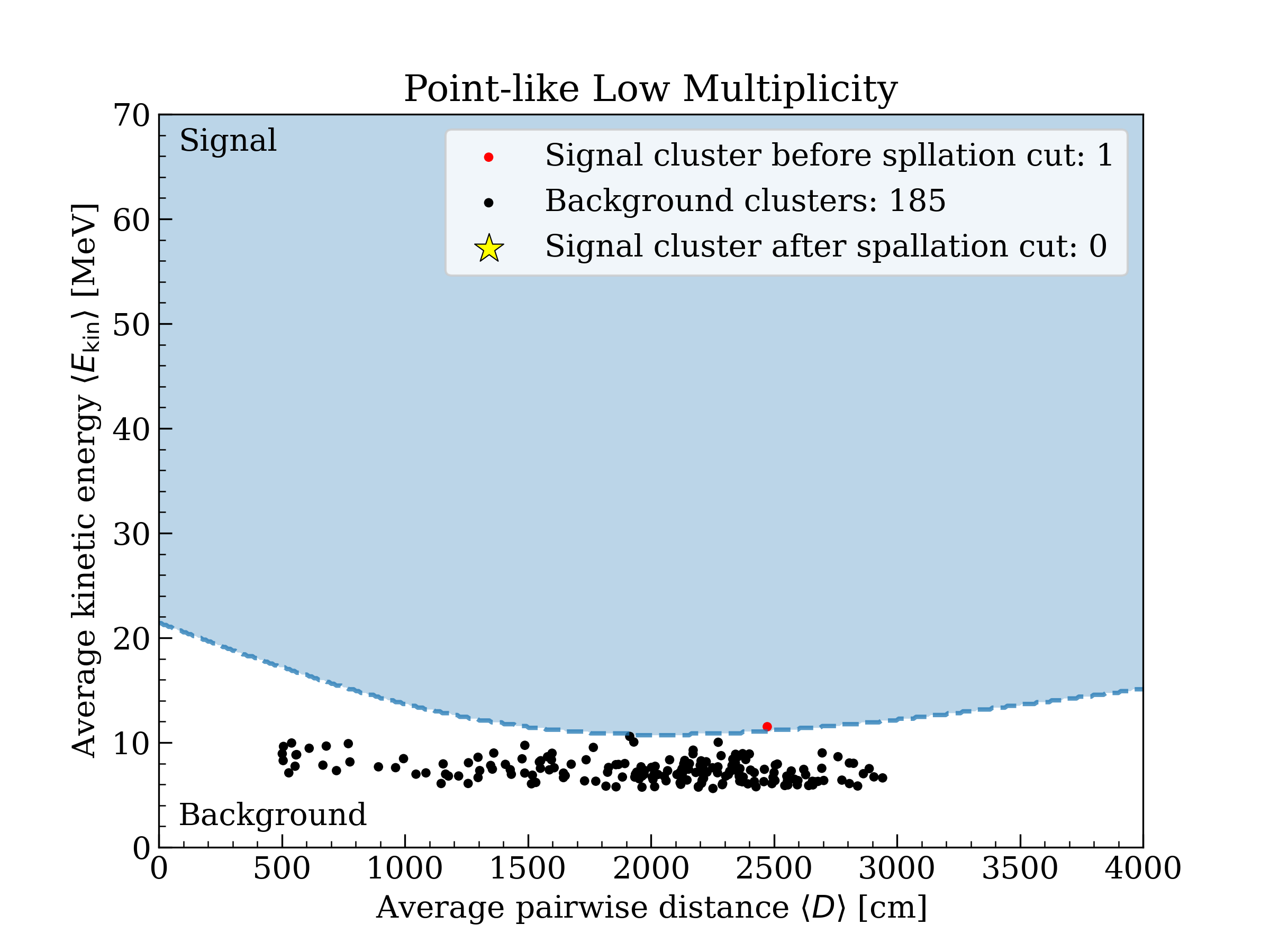}
         \end{minipage}&
         \begin{minipage}{0.5\hsize}
         \centering
         \includegraphics[width=8cm]{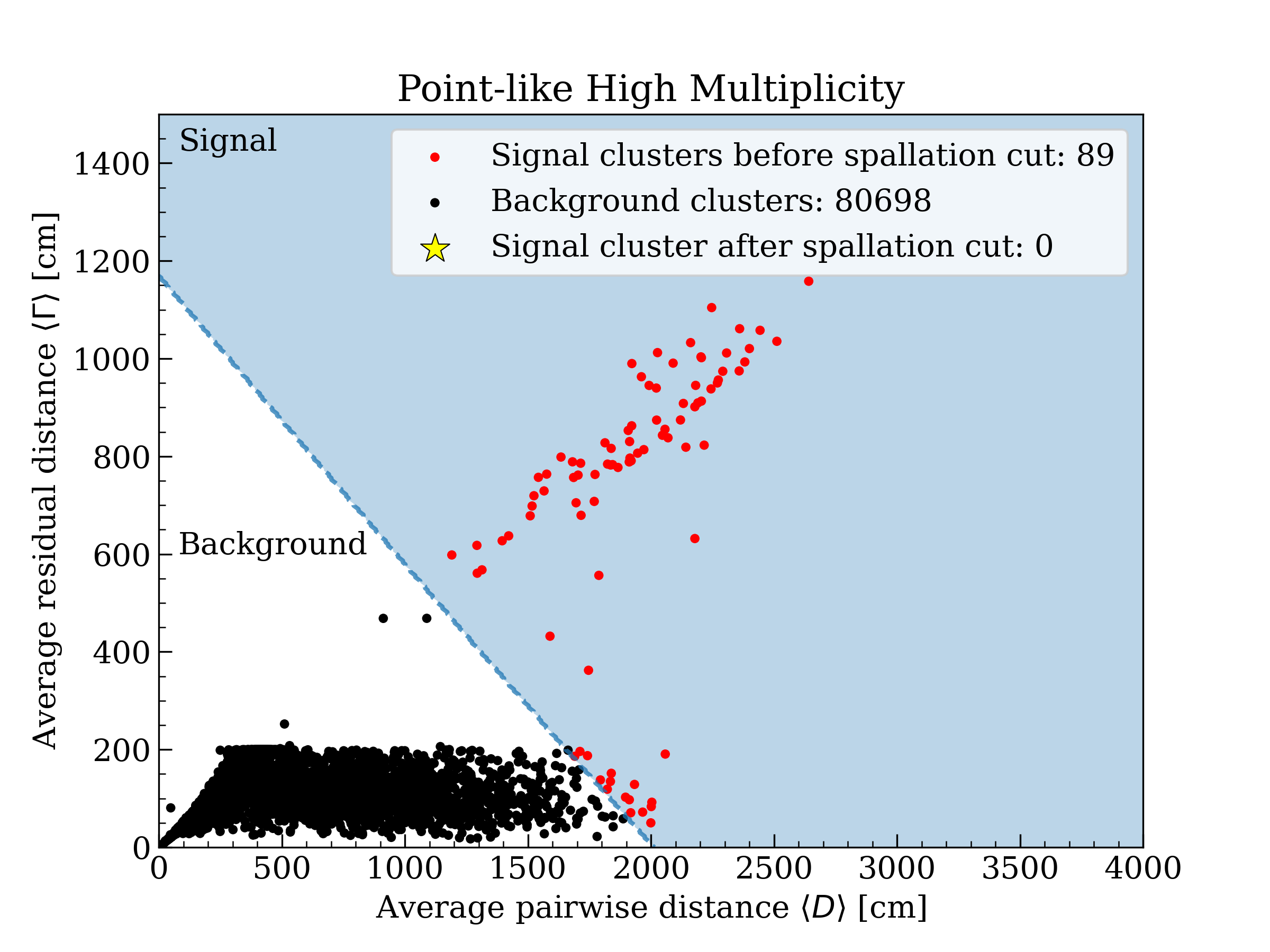}
         \end{minipage} \\
    \end{tabular}
    \caption{Results of the cluster search on the full SK-IV data-set before and after application of the spallation cut.
    Red (black) points denote clusters falling into the blue (white) signal-like (background-like) regions before application of the spallation cut. The star denotes the lone remaining signal-like cluster after application of the spallation cut.}
\label{fig:results}
\end{figure}

\begin{table}[H]
    \centering
    \begin{tabular}{ccc|c}
    \hline \hline
        Category & Signal region & Background region & Total\\
        \hline
         Volume-like       & 2   & 360   & 362 \\
         Plane-like        & 71  & 60901 & 60972\\
         Line-like Low Multiplicity  & 5   & 4852 & 4857\\
         Line-like High Multiplicity  & 93  & 4666 & 4759\\
         Point-like Low Multiplicity & 1   & 185 & 186\\ 
         Point-like High Multiplicity & 89  &  80698 & 80687 \\
         \hline
         Total & 261 &  151662 & 151923\\
         \hline \hline
    \end{tabular}
    \caption{Summary of clusters before the spallation cut.}\label{tab:cluster_1st}
\end{table}

Figure~\ref{fig:results} shows the final remaining clusters before and after application of the spallation cut.
One cluster remains after all of the event selections and it is classified as a line-like high multiplicity cluster.
This cluster is discussed in detail in the next section. 


\section{Discussion}~\label{sec:discussion}

\begin{table}[H]
    \centering
    \begin{tabular}{ccccc}
    \hline \hline
        Time [${\rm s}]$& $x [{\rm\,cm}]$ & $y [{\rm\,cm}]$& $z [{\rm\,cm}]$ & Kinetic energy [$\rm MeV$]\\
        \hline
        0.0 & 414 & 1363 & -1261 & 5.79 \\
        0.83 & -869 & -909 & 1101 & 44.77 \\
        0.95 & -842 &  -845 & 1126 & 9.35 \\
         \hline \hline
    \end{tabular}
    \caption{Event information of the remaining cluster, where time is measured from the first event and $x,y$ and $z$ are vertexes in the tank. The first event was recorded at 19:50:11 on July 7th, 2008 UTC. \label{tab:event_information_in_cluster}}
\end{table}

\begin{figure}
    \centering
    \includegraphics[width=10cm]{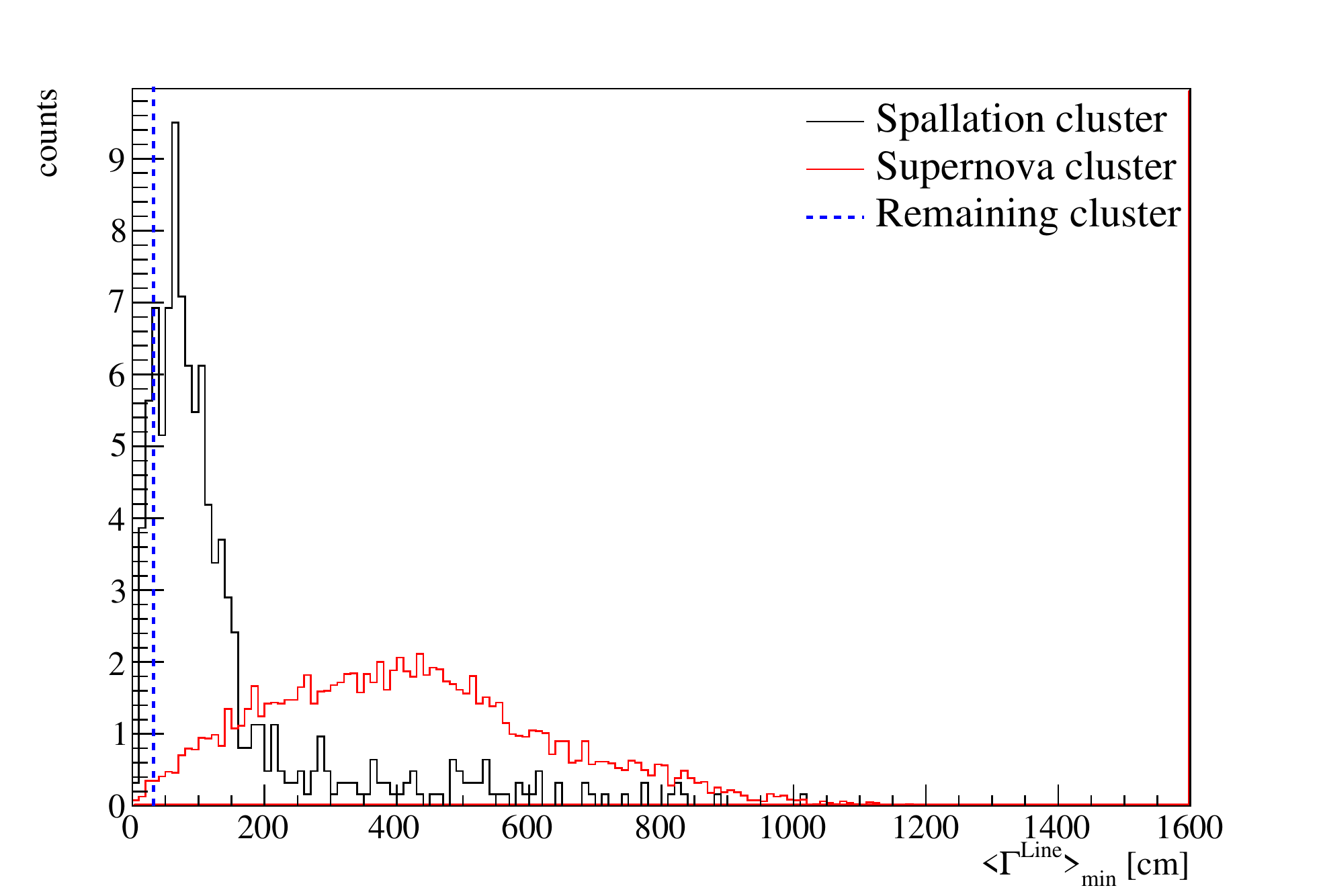}
    \caption{The average residual distance from the best-fit line for line-like clusters. The red histogram shows the distribution for 
supernova clusters and the black is for spallation clusters. The average residual distance of the candidate cluster is 32~cm.}
    \label{fig:rmean_dim_dim1}
\end{figure}

After all analysis cuts one supernova-like cluster remains.
A visual scan of the events in the cluster found that they are consistent with particle interactions and cannot be attributed to noise in the detector or its electronics.
The three events in the remaining cluster have an $\average{E_{\rm kin}}$ of 19.97~MeV and their individual times and vertexes are summarized Table~\ref{tab:event_information_in_cluster}\footnote{The detector coordinates have the origin at the center of the tank with the $x$ and $y$ axes parallel to its top and bottom surfaces. The $z$ axis is along the central axis of the detector. Further details can be found in Ref.~\cite{NAKAHATA1999113}.}.
We define the average residual distance $\average{\Gamma^{\rm Line}}_{\rm min}$, which is the average distance of events from the best-fit line through the cluster:
\begin{equation}
    \average{\Gamma^{\rm Line}}_{\rm min} = \min_{\vec{x}(\alpha)}{\left\{\sqrt{\frac{M}{2(M-2)}\sum_{i=1}^{M}(\vec{d}_i-\vec{x}(\alpha))^2}\right\}},
\end{equation}
where $\vec{x}(\alpha) = \vec{d}_0 + \alpha \vec{n}$, with $\alpha$ a  real number and $n$ a unit vector, specifies a line which passes through the centroid ($\vec{d}_{0}$) of the cluster.
Figure~\ref{fig:rmean_dim_dim1} shows the distribution of this parameter. 
The last two events are separated spatially by only 74~cm while the first is separated from those two by $\sim$3500~cm giving  
$\average D = 2361$~cm. 
Note that if this cluster is attributed to a supernova, the p-value for observing an event with an energy larger than that of the
largest energy event in the cluster is 0.3\% for the Mori model.
Even the most energetic photons from spallation nuclei are below 20~MeV, indicating that this event is 
more consistent with the expected energy of a Michel electron from a decayed muon. 
However, looking at all muons preceding these events none are found with a stopping point consistent with the position of this event.

Another possibility is that light from particles produced in interactions of the form $\nu_{\mu} + \isotope[16]{O} \rightarrow \mu^{-} + \isotope[16]{F}$ or $\bar{\nu}_{\mu} + \isotope[16]{O} \rightarrow \mu^{+} + \isotope[16]{N}$ in coincidence with a 5.79 MeV background event could explain this cluster.  
In the case of the latter, the 44.77~MeV and 9.35~MeV events could be explained by the Michel electron of a below-Cherenkov-threshold $\mu^{+}$ followed by a photon from the  $\isotope[16]{N}$ decay ($\tau=7.13$s), respectively.  For the former, an above-threshold $\mu^{-}$ and its subsequent capture on $\isotope[16]{O}$ to produce $\isotope[16]{N}$ would provide the mechanism. 

Calculating the average residual distance of these events to the best-fit line returned by the line-like fit yields a distance of 32~cm.
Figure~\ref{fig:rmean_dim_dim1} shows the same distribution for selected line-like clusters. 
Supernova clusters peak at around 450\,cm while the peak for spallation clusters is around 70\,cm.
In this metric only 0.59\% (12.4\%) of supernova (spallation) clusters have a residual distance smaller than 32~cm,
indicating this cluster is more consistent with spallation.
Despite the tight residual distance to a line no muons were found passing in the vicinity of the line found by the line-like fit 
in the 60~s of data prior to the start of this cluster. 
For these reasons this cluster of events is not easily explained as either supernova neutrino interactions nor as the decays of spallation products.
As a result, two limit calculations are presented below, 
one of which considers this candidate representative of an unknown background source that occurs once in the current data set and one that does not. 
Note that the candidate cluster is not removed from the observation when computing either limit.



\subsection{Upper limit}
Since no clear candidate supernova clusters were found in the analysis, an upper limit on the rate of supernovae out to 100~kpc is calculated.
This limit is valid for all models and hierarchies considered, except that for the failed supernovae. 
For that case the limit can be extended to 400~kpc assuming the normal mass hierarchy since the detection probability is 1.0 
at this distance, and similarly out to 300~kpc for the inverted hierarchy (c.f. Figure~\ref{fig:mc_predict}).
We have one background cluster and the upper limit at 90\% confidence level is calculated using,
\begin{equation}
    0.9 = \frac{e^{bT_{\rm live}}}{bT_{\rm live}+1}\int^{\lambda_{\rm lim}}_{0}e^{-(\lambda^\prime+bT_{\rm live})}(\lambda^\prime + bT_{\rm live})d\lambda^\prime,
\end{equation}
where $b$ is background rate and $T_{\rm live}$ is the livetime.

The upper limit on the supernova rate is then given by: 
\begin{equation}
    R_{\rm SN} = \frac{\lambda^{\rm lim}}{T_{\rm live}p_{\rm detect}},
\end{equation}
where $p_{\rm detect}$ is the detection probability at 100\,kpc.
The background rate $b$ is estimated from 204.8 days of data taken at random from normal runs in SK-IV after applying the spallation cut.
We increase the effective statistics by generating 100 additional clusters for each passing cluster in the same manner as used in the ML training described above. 
As a result, 27 clusters are found across all signal regions.
The background rate is therefore
\begin{equation}
    b = \frac{27}{204.8 {\rm\, days} \times 100} = 0.4812{\rm\, yr^{-1}}.
\end{equation}
Tests with larger numbers of additional clusters yielded nearly identical background estimations.
However, this estimation may not account for the type of cluster responsible for the remaining candidate in Figure~\ref{fig:results}.
To account for this possibility, we additionally consider a modified background rate, where this candidate is considered to represent an unmodelled background,
\begin{equation}
    b^\prime = 0.4812 + \frac{1}{3318.41 {\rm\,days}}\times 365 = 0.5911 {\rm\,yr^{-1}}.
\end{equation}
Using either background estimate the resulting upper limit at 90\% C.L. is unchanged in the first two decimal places. The resulting upper limit from this analysis is 0.29 supernovae per year out to 100~kpc.

\section{Summary}\label{sec:summary}
This analysis presents a search for neutrinos from distant supernovae carried out using roughly ten years of Super-Kamiokande data. 
Relative to the previous analysis from SK, the present work adopts modern supernova simulations and event selections optimized 
with machine learning tools. 
For the Mori model the detection probability is 98.5\% at 100~kpc and 39.1\% at 200~kpc assuming the normal mass hierarchy. 
Using the Nakazato model the probabilities are 100\% and 84.2\%, respectively.
No significant signal excess was found in the search, resulting in a 90\% confidence level limit on the rate of supernovae of 0.29 per year out to 100~kpc for these models and out to 300~kpc for the failed supernova model.
These may be compared with recent result from LVD of 0.09 per year out to 25~kpc~\citep{Vigorito:2020vjj} and to 0.32 per year from our previous search~\citep{Ikeda:2007sa}.  
Despite the decreased event rate expected from modern models relative to that in the latter, the present work slightly improves on its limit at the same distances.

\appendix
\section{Performance of the goodness parameter}\label{sec:perfomance_ovaq}
\begin{figure}
    \centering
\begin{tabular}{cc}
         \begin{minipage}{0.5\hsize}
         \centering
         \includegraphics[width=10cm]{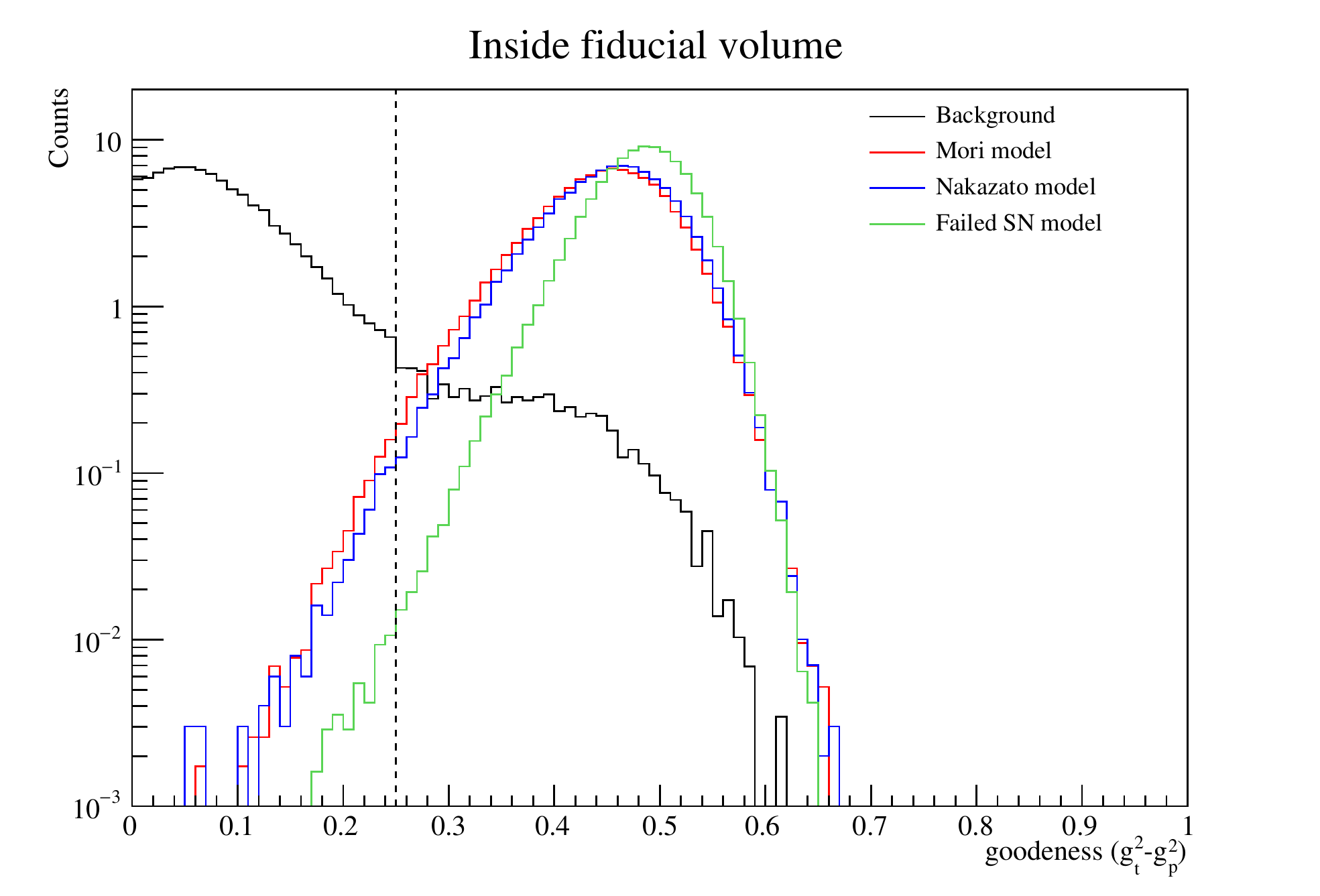}
         \end{minipage}&
         \begin{minipage}{0.5\hsize} 
         \centering
         \includegraphics[width=10cm]{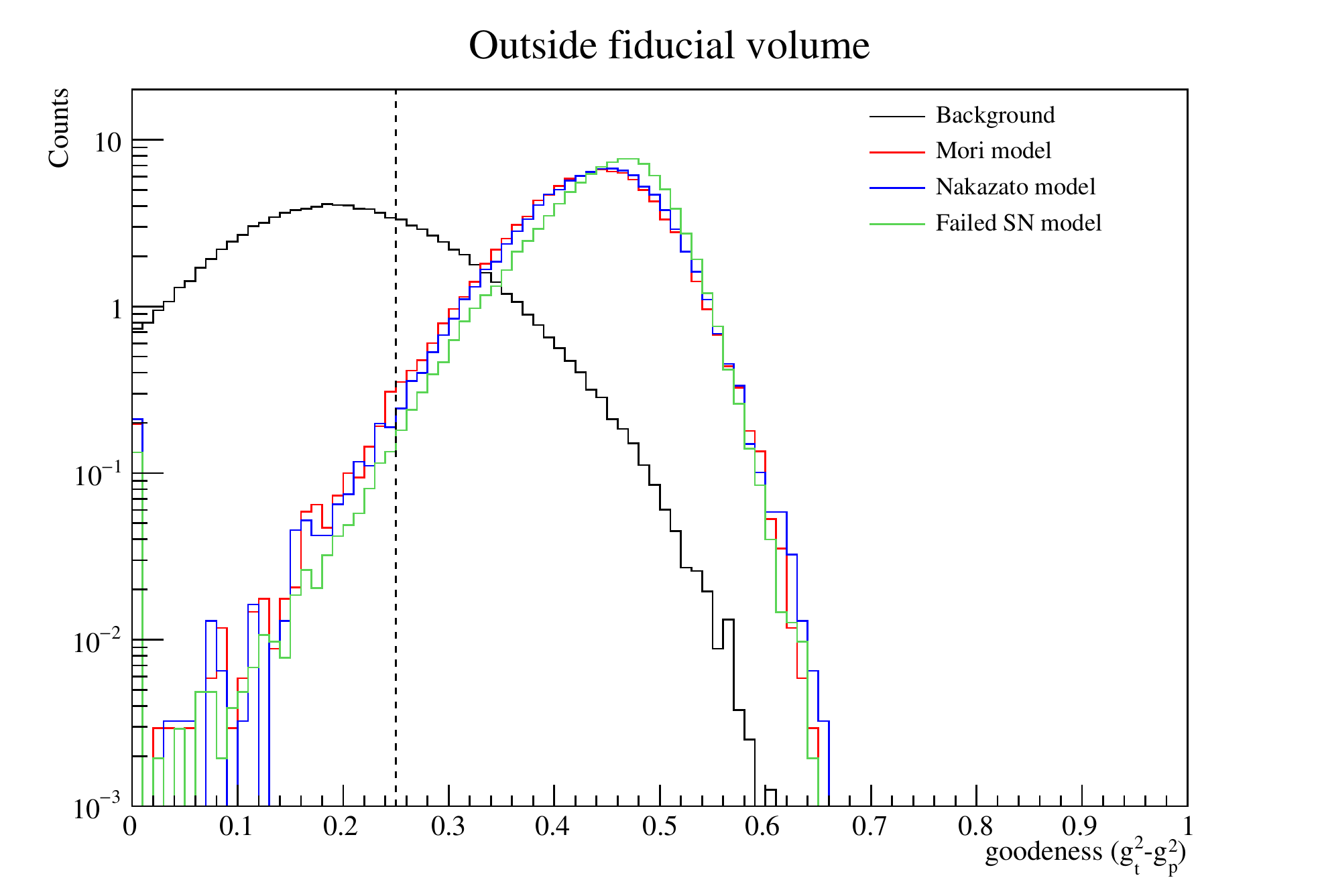}
         \end{minipage} \\
         \end{tabular}
         \caption{Goodness distribution of background and supernova events. The left (right) panel shows events reconstructed inside (outside) of the fiducial volume and the right panel shows those outside. The dashed lines indicate are at the cut position, $g_t^2 - g_p^2 = 0.25$. All histograms are normalized to an area of 100 events and events are required to have energies greater than 5.0 MeV.\label{fig:ovaq_plot}}
\end{figure}

This Appendix details the performance of the goodness parameter $g^2_t - g^2_p$ described in Equations~\ref{eq:g_t} and~\ref{eq:g_p} for supernova neutrino events and background events.
Figure~\ref{fig:ovaq_plot} shows the goodness distribution of events both inside and outside of the fiducial volume. 
The background distribution in either case largely populates the region below 0.25 whereas signal events are above this threshold. 
As a result, the present analysis uses events with goodness greater than 0.25.
Table~\ref{tab:summary_ovaq} summarizes the fraction of events above this threshold for the background and signal models.

\begin{table}[]
    \centering
    \begin{tabular}{ccccc}
    \hline\hline
         &  Background & Mori & Nakazato & Failed SN \\ \hline
        Inside & 5.0\% & 94.8\% & 94.9\% & 96.0\% \\
        Outside & 29.5\% & 66.7\% & 62.5\% & 64.5\% \\
        \hline\hline
    \end{tabular}
    \caption{Fraction of events passing the goodness threshold of 0.25 for events reconstructed inside (Inside) and outside (Outside) of the fiducial volume.    \label{tab:summary_ovaq}}

\end{table}

\bibliography{iopart-num.bib}

\begin{thebibliography}{}
\expandafter\ifx\csname natexlab\endcsname\relax\def\natexlab#1{#1}\fi
\providecommand{\url}[1]{\href{#1}{#1}}
\providecommand{\dodoi}[1]{doi:~\href{http://doi.org/#1}{\nolinkurl{#1}}}
\providecommand{\doeprint}[1]{\href{http://ascl.net/#1}{\nolinkurl{http://ascl.net/#1}}}
\providecommand{\doarXiv}[1]{\href{https://arxiv.org/abs/#1}{\nolinkurl{https://arxiv.org/abs/#1}}}

\bibitem[{Abe {et~al.}(2016{\natexlab{a}})}]{Abe:2016waf}
Abe, K., {et~al.} 2016{\natexlab{a}}, Astropart. Phys., 81, 39,
  \dodoi{10.1016/j.astropartphys.2016.04.003}

\bibitem[{Abe {et~al.}(2016{\natexlab{b}})}]{Abe:2016nxk}
---. 2016{\natexlab{b}}, Phys. Rev. D, 94, 052010,
  \dodoi{10.1103/PhysRevD.94.052010}

\bibitem[{Abe {et~al.}(2022)}]{Super-Kamiokande:2021the}
---. 2022, Nucl. Instrum. Meth. A, 1027, 166248,
  \dodoi{10.1016/j.nima.2021.166248}

\bibitem[{Aglietta {et~al.}(1988)Aglietta, Badino, Bologna, Castagnoli,
  Castellina, Dadykin, Fulgione, Galeotti, Kalchukov, Kortchaguin, Kortchaguin,
  Malguin, Ryassny, Ryazhkaya, Saavedra, Talochkin, Trinchero, Vernetto,
  Zatsepin, \& Yakushev}]{AGLIETTA1988453}
Aglietta, M., Badino, G., Bologna, G., {et~al.} 1988, Nuclear Physics B -
  Proceedings Supplements, 3, 453,
  \dodoi{https://doi.org/10.1016/0920-5632(88)90196-X}

\bibitem[{Aharmim {et~al.}(2011)Aharmim, Ahmed, Anthony, Barros, Beier,
  Bellerive, Beltran, Bergevin, Biller, Boudjemline, Boulay, Cai, Chan,
  Chauhan, Chen, Cleveland, Dai, Deng, Detwiler, Doucas, Drouin, Duncan,
  Dunford, Earle, Elliott, Evans, Ewan, Farine, Fergani, Fleurot, Ford,
  Formaggio, Gagnon, Goon, Graham, Guillian, Habib, Hahn, Hallin, Hallman,
  Harvey, Hazama, Heintzelman, Heise, Helmer, Hime, Howard, Huang, Jamieson,
  Jelley, Jerkins, Klein, Kos, Kraus, Krauss, Kutter, Kyba, Law, Lesko, Leslie,
  Loach, MacLellan, Majerus, Mak, Maneira, Martin, McCauley, McDonald, Miller,
  Monreal, Monroe, Nickel, Noble, O{\textquotesingle}Keeffe, Oblath, Gann,
  Oser, Ott, Peeters, Poon, Prior, Reitzner, Rielage, Robertson, Robertson,
  Schwendener, Secrest, Seibert, Simard, Simpson, Sinclair, Skensved, Sonley,
  Stonehill, Te{\v{s}}i{\'{c}}, Tolich, Tsui, Berg, VanDevender, Virtue,
  Tseung, Watson, West, Wilkerson, Wilson, Wright, Yeh, Zhang, , \&
  and}]{Aharmim_2011}
Aharmim, B., Ahmed, S.~N., Anthony, A.~E., {et~al.} 2011, The Astrophysical
  Journal, 728, 83, \dodoi{10.1088/0004-637x/728/2/83}

\bibitem[{Alexeyev {et~al.}(1988)Alexeyev, Alexeyeva, Krivosheina, \&
  Volchenko}]{ALEXEYEV1988209}
Alexeyev, E., Alexeyeva, L., Krivosheina, I., \& Volchenko, V. 1988, Physics
  Letters B, 205, 209, \dodoi{https://doi.org/10.1016/0370-2693(88)91651-6}

\bibitem[{Ambrosio {et~al.}(1998)}]{MACRO:1997enu}
Ambrosio, M., {et~al.} 1998, Astropart. Phys., 8, 123,
  \dodoi{10.1016/S0927-6505(97)00032-7}

\bibitem[{Ando {et~al.}(2005)Ando, Beacom, \& Y\"uksel}]{PhysRevLett.95.171101}
Ando, S., Beacom, J.~F., \& Y\"uksel, H. 2005, Phys. Rev. Lett., 95, 171101,
  \dodoi{10.1103/PhysRevLett.95.171101}

\bibitem[{Asakura {et~al.}(2016)}]{Asakura:2015bga}
Asakura, K., {et~al.} 2016, Astrophys. J., 818, 91,
  \dodoi{10.3847/0004-637X/818/1/91}

\bibitem[{Bionta {et~al.}(1987)Bionta, Blewitt, Bratton, Casper, Ciocio, Claus,
  Cortez, Crouch, Dye, Errede, Foster, Gajewski, Ganezer, Goldhaber, Haines,
  Jones, Kielczewska, Kropp, Learned, LoSecco, Matthews, Miller, Mudan, Park,
  Price, Reines, Schultz, Seidel, Shumard, Sinclair, Sobel, Stone, Sulak,
  Svoboda, Thornton, van~der Velde, \& Wuest}]{PhysRevLett.58.1494}
Bionta, R.~M., Blewitt, G., Bratton, C.~B., {et~al.} 1987, Phys. Rev. Lett.,
  58, 1494, \dodoi{10.1103/PhysRevLett.58.1494}

\bibitem[{Bishop(2006)}]{10.5555/1162264}
Bishop, C.~M. 2006, Pattern Recognition and Machine Learning (Information
  Science and Statistics) (Berlin, Heidelberg: Springer-Verlag)

\bibitem[{Bruno {et~al.}(2017)Bruno, Molinario, Fulgione, \&
  Vigorito}]{Bruno:2017oxg}
Bruno, G., Molinario, A., Fulgione, W., \& Vigorito, C. 2017, J. Phys. Conf.
  Ser., 888, 012256, \dodoi{10.1088/1742-6596/888/1/012256}

\bibitem[{Cravens {et~al.}(2008)}]{Cravens:2008aa}
Cravens, J., {et~al.} 2008, Phys. Rev. D, 78, 032002,
  \dodoi{10.1103/PhysRevD.78.032002}

\bibitem[{Dighe \& Smirnov(2000)}]{Dighe:1999bi}
Dighe, A.~S., \& Smirnov, A.~Y. 2000, Phys. Rev. D, 62, 033007,
  \dodoi{10.1103/PhysRevD.62.033007}

\bibitem[{Fukuda {et~al.}(2003)}]{Fukuda:2002uc}
Fukuda, Y., {et~al.} 2003, Nucl. Instrum. Meth. A, 501, 418,
  \dodoi{10.1016/S0168-9002(03)00425-X}

\bibitem[{Hirata {et~al.}(1987)Hirata, Kajita, Koshiba, Nakahata, Oyama, Sato,
  Suzuki, Takita, Totsuka, Kifune, Suda, Takahashi, Tanimori, Miyano, Yamada,
  Beier, Feldscher, Kim, Mann, Newcomer, Van, Zhang, \&
  Cortez}]{PhysRevLett.58.1490}
Hirata, K., Kajita, T., Koshiba, M., {et~al.} 1987, Phys. Rev. Lett., 58, 1490,
  \dodoi{10.1103/PhysRevLett.58.1490}

\bibitem[{Hosaka {et~al.}(2006)}]{Hosaka:2005um}
Hosaka, J., {et~al.} 2006, Phys. Rev. D, 73, 112001,
  \dodoi{10.1103/PhysRevD.73.112001}

\bibitem[{Ikeda {et~al.}(2007)}]{Ikeda:2007sa}
Ikeda, M., {et~al.} 2007, Astrophys. J., 669, 519, \dodoi{10.1086/521547}

\bibitem[{K\"opke(2018)}]{Kopke:2017req}
K\"opke, L. 2018, J. Phys. Conf. Ser., 1029, 012001,
  \dodoi{10.1088/1742-6596/1029/1/012001}

\bibitem[{Li \& Beacom(2014)}]{Li:2014sea}
Li, S.~W., \& Beacom, J.~F. 2014, Phys. Rev. C, 89, 045801,
  \dodoi{10.1103/PhysRevC.89.045801}

\bibitem[{Li \& Beacom(2015)}]{Li:2015lxa}
---. 2015, Phys. Rev. D, 92, 105033, \dodoi{10.1103/PhysRevD.92.105033}

\bibitem[{Monzani(2006)}]{Monzani:2006jg}
Monzani, M.~E. 2006, Nuovo Cim. C, 29, 269, \dodoi{10.1393/ncc/i2005-10230-2}

\bibitem[{Mori {et~al.}(2020)Mori, Suwa, Nakazato, Sumiyoshi, Harada, Harada,
  Koshio, \& Wendell}]{10.1093/ptep/ptaa185}
Mori, M., Suwa, Y., Nakazato, K., {et~al.} 2020, Progress of Theoretical and
  Experimental Physics, 2021, \dodoi{10.1093/ptep/ptaa185}

\bibitem[{Nakahata {et~al.}(1999)Nakahata, Fukuda, Hayakawa, Ichihara, Inoue,
  Ishihara, Ishino, Itow, Kajita, Kameda, Kasuga, Kobayashi, Kobayashi, Koshio,
  Martens, Miura, Nakayama, Okada, Okumura, Sakurai, Shiozawa, Suzuki,
  Takeuchi, Totsuka, Yamada, Earl, Habig, Kearns, Messier, Scholberg, Stone,
  Sulak, Walter, Goldhaber, Barszczak, Casper, Gajewski, Halverson, Hsu, Kropp,
  Price, Reines, Smy, Sobel, Vagins, Ganezer, Keig, Ellsworth, Tasaka,
  Flanagan, Kibayashi, Learned, Matsuno, Stenger, Takemori, Ishii, Kanzaki,
  Kobayashi, Mine, Nakamura, Nishikawa, Oyama, Sakai, Sakuda, Sasaki, Echigo,
  Kohama, Suzuki, Haines, Blaufuss, Kim, Sanford, Svoboda, Chen, Conner,
  Goodman, Sullivan, Hill, Jung, Mauger, McGrew, Sharkey, Viren, Yanagisawa,
  Doki, Miyano, Okazawa, Saji, Takahata, Nagashima, Takita, Yamaguchi, Yoshida,
  Kim, Etoh, Fujita, Hasegawa, Hasegawa, Hatakeyama, Iwamoto, Koga, Maruyama,
  Ogawa, Shirai, Suzuki, Tsushima, Koshiba, Nemoto, Nishijima, Futagami,
  Hayato, Kanaya, Kaneyuki, Watanabe, Kielczewska, Doyle, George, Stachyra,
  Wai, Wilkes, Young, \& Kobayashi}]{NAKAHATA1999113}
Nakahata, M., Fukuda, Y., Hayakawa, T., {et~al.} 1999, Nuclear Instruments and
  Methods in Physics Research Section A: Accelerators, Spectrometers, Detectors
  and Associated Equipment, 421, 113,
  \dodoi{https://doi.org/10.1016/S0168-9002(98)01200-5}

\bibitem[{Nakazato {et~al.}(2013)Nakazato, Sumiyoshi, Suzuki, Totani, Umeda, \&
  Yamada}]{Nakazato:2012qf}
Nakazato, K., Sumiyoshi, K., Suzuki, H., {et~al.} 2013, Astrophys. J. Suppl.,
  205, 2, \dodoi{10.1088/0067-0049/205/1/2}

\bibitem[{Nishino {et~al.}(2009)Nishino, Awai, Hayato, Nakayama, Okumura,
  Shiozawa, Takeda, Ishikawa, Minegishi, \& Arai}]{Nishino:2009zu}
Nishino, H., Awai, K., Hayato, Y., {et~al.} 2009, Nucl. Instrum. Meth. A, 610,
  710, \dodoi{10.1016/j.nima.2009.09.026}

\bibitem[{Pedregosa {et~al.}(2011)Pedregosa, Varoquaux, Gramfort, Michel,
  Thirion, Grisel, Blondel, Prettenhofer, Weiss, Dubourg, Vanderplas, Passos,
  Cournapeau, Brucher, Perrot, \& {{\'E}}douard
  Duchesnay}]{JMLR:v12:pedregosa11a}
Pedregosa, F., Varoquaux, G., Gramfort, A., {et~al.} 2011, Journal of Machine
  Learning Research, 12, 2825.
\newblock \url{http://jmlr.org/papers/v12/pedregosa11a.html}

\bibitem[{Rozwadowska {et~al.}(2021)Rozwadowska, Vissani, \&
  Cappellaro}]{Rozwadowska:2021lll}
Rozwadowska, K., Vissani, F., \& Cappellaro, E. 2021, New Astron., 83, 101498,
  \dodoi{10.1016/j.newast.2020.101498}

\bibitem[{Sato \& Suzuki(1987)}]{PhysRevLett.58.2722}
Sato, K., \& Suzuki, H. 1987, Phys. Rev. Lett., 58, 2722,
  \dodoi{10.1103/PhysRevLett.58.2722}

\bibitem[{{Totani} {et~al.}(1998){Totani}, {Sato}, {Dalhed}, \&
  {Wilson}}]{totani}
{Totani}, T., {Sato}, K., {Dalhed}, H.~E., \& {Wilson}, J.~R. 1998, \apj, 496,
  216, \dodoi{10.1086/305364}

\bibitem[{Vigorito {et~al.}(2020)Vigorito, Bruno, Fulgione, \&
  Molinario}]{Vigorito:2020vjj}
Vigorito, C.~F., Bruno, G., Fulgione, W., \& Molinario, A. 2020, PoS, ICRC2019,
  1028, \dodoi{10.22323/1.358.1028}

\bibitem[{Wright {et~al.}(2016)Wright, Nagaraj, Kneller, Scholberg, \&
  Seitenzahl}]{Wright_2016}
Wright, W.~P., Nagaraj, G., Kneller, J.~P., Scholberg, K., \& Seitenzahl, I.~R.
  2016, Physical Review D, 94, \dodoi{10.1103/physrevd.94.025026}

\bibitem[{Yamada {et~al.}(2010)Yamada, Awai, Hayato, Kaneyuki, Kouzuma,
  Nakayama, Nishino, Okumura, Obayashi, Shimizu, Shiozawa, Takeda, Heng, Yang,
  Chen, Tanaka, Yokozawa, Koshio, Moriyama, Arai, Ishikawa, Minegishi, \&
  Uchida}]{5446533}
Yamada, S., Awai, K., Hayato, Y., {et~al.} 2010, IEEE Transactions on Nuclear
  Science, 57, 428, \dodoi{10.1109/TNS.2009.2034854}

\bibitem[{Zhang {et~al.}(2016)}]{Super-Kamiokande:2015xra}
Zhang, Y., {et~al.} 2016, Phys. Rev. D, 93, 012004,
  \dodoi{10.1103/PhysRevD.93.012004}

\bibitem[{Zuber(2015)}]{Zuber:2015ita}
Zuber, K. 2015, Nucl. Part. Phys. Proc., 265-266, 233,
  \dodoi{10.1016/j.nuclphysbps.2015.06.059}

\end{thebibliography}


\end{document}